%% file: ANA-HION-2019-08-PAPER.tex
\documentclass[cernpreprint,atlasdraft=false,texlive=2016,UKenglish,texmf]{atlasdoc}
 
\usepackage{atlaspackage}
\usepackage{atlasheavyion}
\usepackage{lipsum}
\usepackage{todonotes}

\usepackage{atlasbiblatex}
 
\usepackage{atlascontribute}
 
\usepackage{atlasphysics}

\graphicspath{{logos/}{figures/}}
 
\usepackage{ANA-HION-2019-08-PAPER-defs}

\AtlasTitle{Measurement of light-by-light scattering and search for
axion-like particles  with $2.2\;\mathrm{nb^{-1}}$ of Pb+Pb data with the ATLAS detector}

\AtlasAbstract{
This paper describes a measurement of light-by-light scattering based on Pb+Pb collision data recorded by the ATLAS experiment during Run 2 of the LHC.
The study uses $2.2$~nb$^{-1}$  of integrated luminosity collected in 2015 and 2018 at $\sqrt{s_\mathrm{NN}}=5.02$~TeV.
Light-by-light scattering candidates are selected in events with two
photons produced exclusively, each
with transverse energy $E_{\textrm{T}}^{\gamma} > 2.5$~GeV, pseudorapidity $|\eta_{\gamma}| < 2.37$, diphoton invariant mass $m_{\gamma\gamma} > 5$~GeV, and with small diphoton transverse momentum and diphoton acoplanarity.
The integrated and differential fiducial cross sections are measured and compared with theoretical predictions.
The diphoton invariant mass distribution is used to set limits on the
production of axion-like particles. This result provides the most stringent
limits to date on axion-like particle production for masses in the range
6--100~GeV. Cross sections above 2 to 70~nb are excluded at the 95\%
CL in that mass interval.}
 
\author{The ATLAS Collaboration}

\PreprintIdNumber{CERN-EP-2020-135}

\AtlasJournalRef{JHEP 03 (2021) 243}
\AtlasDOI{10.1007/JHEP03(2021)243}
\hypersetup{pdftitle={ATLAS document},pdfauthor={The ATLAS Collaboration}}
\usepackage{textpos}
\clearpage
 
\pdfoutput=1
\begin{document}
 
\maketitle
 
\clearpage
\newpage
 
\tableofcontents

\section{\label{sec:Intro}Introduction}
 
\newcommand{\AtlasCoordFootnote}{
ATLAS uses a right-handed coordinate system with origin at the nominal interaction point in the centre of the detector and the $z$-axis along the beam pipe.
The $x$-axis points from the IP to the centre of the LHC ring,
and the $y$-axis points upwards.
Cylindrical coordinates $(r,\phi)$ are used in the transverse plane,
$\phi$ being the azimuthal angle around the $z$-axis.
The pseudorapidity is defined in terms of the polar angle $\theta$ as $\eta = -\ln \tan(\theta/2)$.
Angular distance is measured in units of $\Delta R \equiv \sqrt{(\Delta\eta)^{2} + (\Delta\phi)^{2}}$.
The transverse energy of a photon or electron is $\et= E/\cosh(\eta)$, where $E$ is its energy.}
 
\begin{figure}[b!]
\centering
\includegraphics[width=0.45\textwidth]{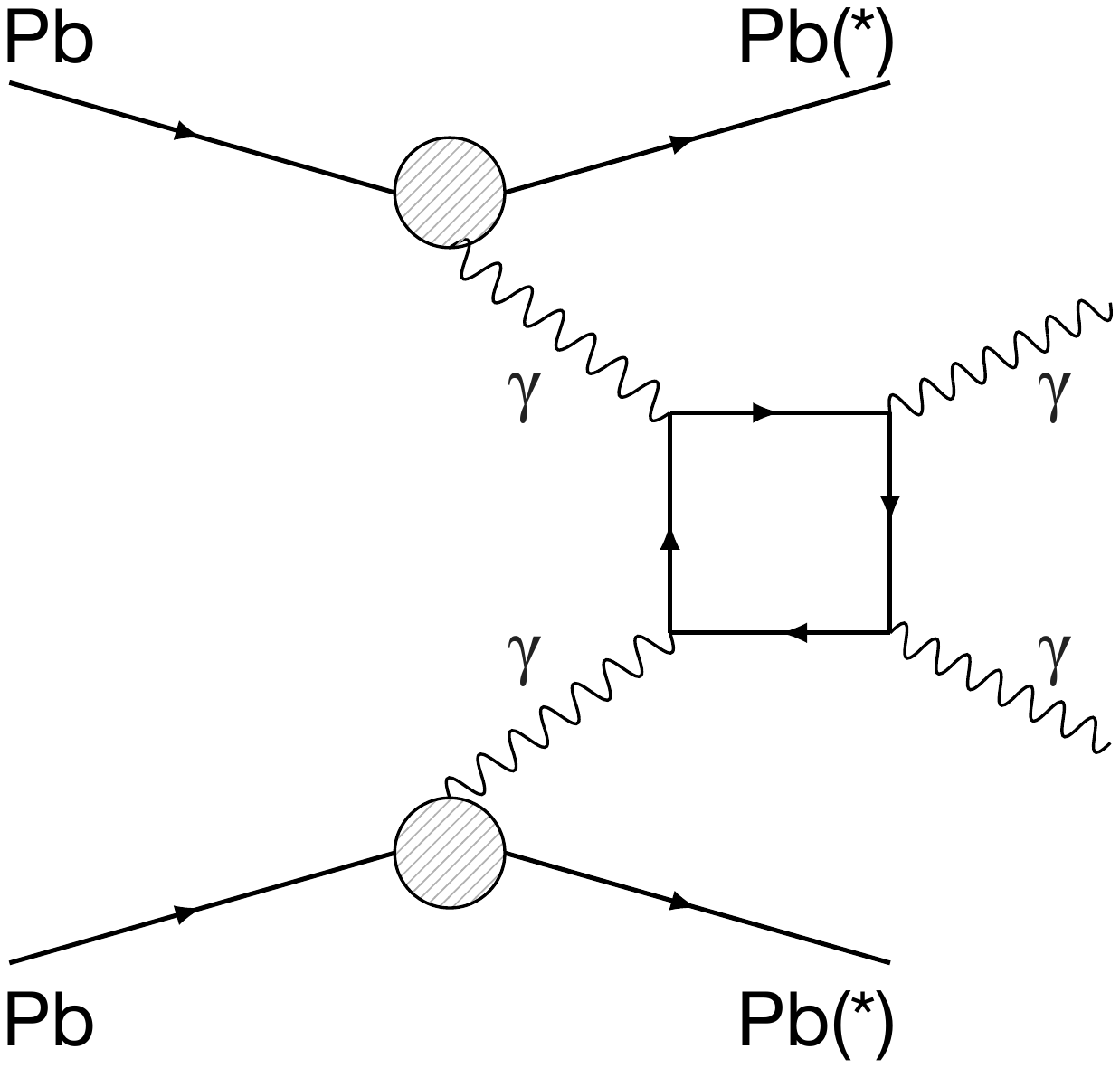}
\includegraphics[width=0.45\textwidth]{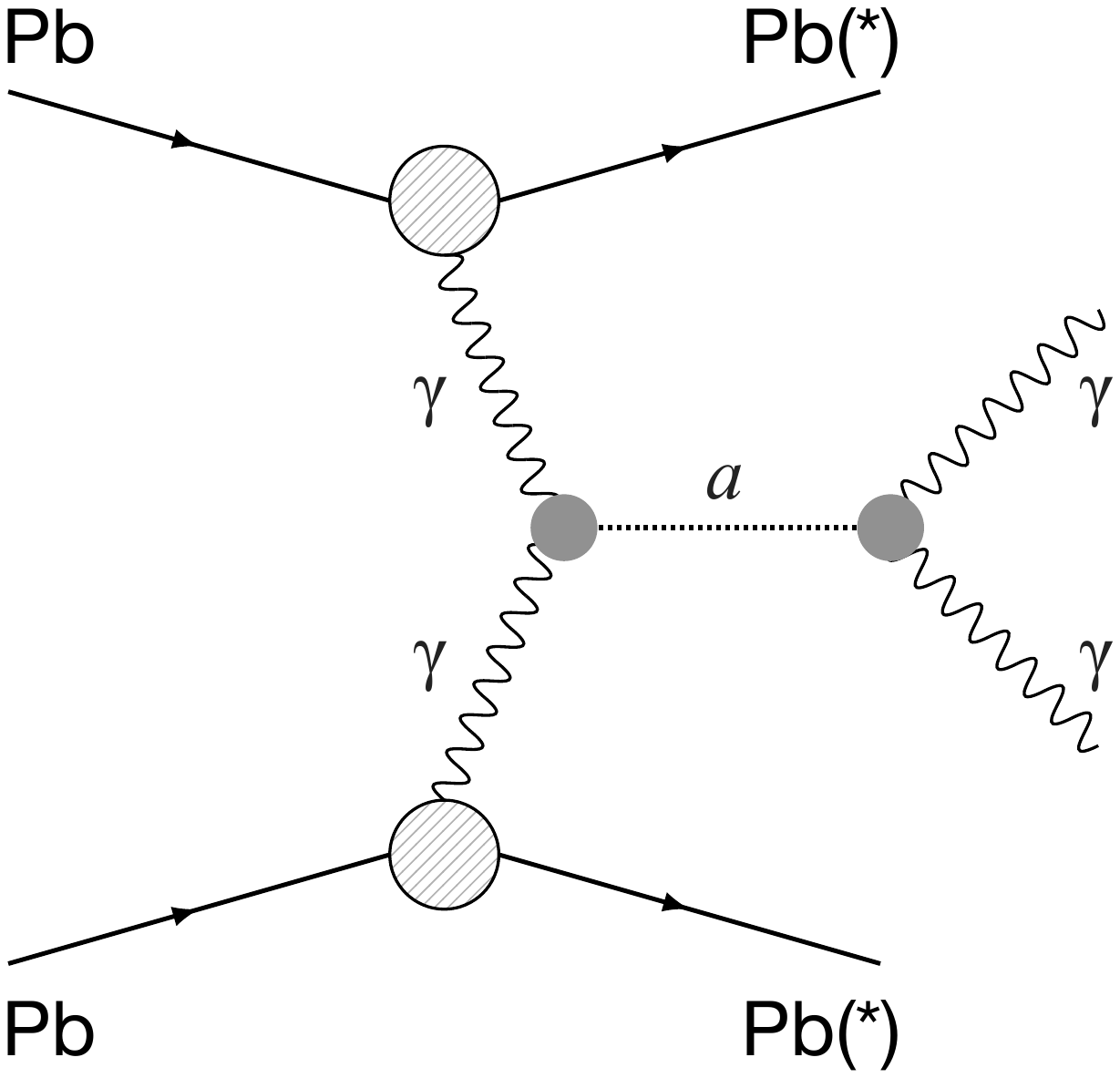}
\caption{\label{fig:UPC}Schematic diagrams of (left) SM LbyL scattering and (right) axion-like particle production in \PbPb\ UPC.
A potential electromagnetic excitation of the outgoing Pb ions is denoted by $(^{\ast})$.}
\end{figure}

Light-by-light~(LbyL) scattering, $\gamma\gamma\rightarrow\gamma\gamma$, is a process in the Standard Model~(SM) that proceeds at lowest order in quantum electrodynamics~(QED) via virtual one-loop box diagrams involving charged fermions (leptons and quarks) and $W^{\pm}$ bosons~(Figure~\ref{fig:UPC}). LbyL interactions can occur in relativistic heavy-ion collisions at any impact parameters. However, the large impact parameters i.e. larger than twice the radius of the ions, are experimentally preferred as the strong interaction does not play a role in these ultra-peripheral collision~(UPC) events.
In general, UPC events allow studies of processes involving nuclear photoexcitation, photoproduction of hadrons, and two-photon interactions. Comprehensive reviews of UPC physics can be found in Refs.~\cite{Baltz:2007kq, Klein:2020fmr}.
The electromagnetic (EM) fields produced by the colliding Pb nuclei
can be treated as a beam of quasi-real photons with a small virtuality of $Q^2 < 1/R^2$, where $R$ is the radius of the nuclear charge distribution and so $Q^2 < 10^{-3}~\GeV^2$~\cite{Fermi:1925fq,vonWeizsacker:1934nji,PhysRev.45.729}.
The cross section for the reaction $\textrm{Pb+Pb}\,(\gamma\gamma)\rightarrow \textrm{Pb}^{(\ast)}\textrm{+}\textrm{Pb}^{(\ast)}\,\gamma\gamma$ can then be calculated by convolving the respective photon flux with the elementary cross section for the process $\gamma\gamma\rightarrow \gamma\gamma$, with a possible EM excitation~\cite{ALICE:2012aa}, denoted by~$(^{\ast})$.
Since the photon flux associated with each nucleus scales as $Z^2$, the LbyL cross section is strongly enhanced relative to proton--proton ($pp$) collisions.
 
In this measurement, the final-state signature of interest is the exclusive production of two photons,  where the diphoton final state is measured in the detector surrounding the Pb+Pb interaction region, and the incoming Pb ions survive the EM interaction.
Hence, one expects that two low-energy photons will be detected with no further activity in the central detector. In particular, no reconstructed charged-particle tracks originating from the Pb+Pb interaction point are expected.

The LbyL process has been proposed as a  sensitive channel to study physics beyond the SM.
Modifications of the $\gamma\gamma\rightarrow\gamma\gamma$  scattering rates can be induced by new exotic charged particles~\cite{Fichet:2014uka}
and by the presence of extra spatial dimensions~\cite{Inan:2019ugz}.
The LbyL cross sections are also sensitive to Born--Infeld extensions of QED~\cite{Ellis:2017edi},
Lorentz-violating operators in electrodynamics~\cite{Kostelecky:2018yfa},
and the presence of space-time non-commutativity in QED~\cite{Horvat:2020ycy}.
Additionally, new neutral particles, such as axion-like particles~(ALP), can also contribute in the form of
narrow diphoton resonances~\cite{Knapen:2016moh}, as shown in Figure~\ref{fig:UPC}.
ALPs are relatively light, gauge-singlet (pseudo-)scalar particles that appear in many theories with a spontaneously broken global
symmetry. Their masses and couplings to SM particles may range over many orders of
magnitude.
The previous ATLAS searches involving ALP decays to photons are based on $pp$ collision data~\cite{EXOT-2013-24,EXOT-2017-08}.

LbyL scattering via an electron loop has been precisely, albeit indirectly, tested in measurements of the anomalous magnetic moment of the electron and muon~\cite{Hanneke:2008tm, Bennett:2006fi}.
The $\gamma\gamma\rightarrow \gamma\gamma$ reaction has been measured in photon scattering in the Coulomb field of a nucleus (Delbrück scattering)~\cite{Wilson:1953zz,PhysRevD.8.3813,Schumacher:1975kv,PhysRevC.58.2844} and in the photon splitting process~\cite{Akhmadaliev:2001ik}.
A related process, in which initial photons fuse to form a pseudoscalar meson which subsequently decays into a pair of photons, has been studied at electron--positron colliders~\cite{Bartel:1985zw, Aihara:1985uc, Williams:1988sg}.
 
The authors of Ref.~\cite{Enterria:2013yra} proposed to measure LbyL scattering by exploiting the large photon fluxes available in heavy-ion collisions at the LHC.
The first direct evidence of the LbyL process in \PbPb UPC at the LHC was established by the  ATLAS~\cite{Aaboud:2017bwk} and CMS~\cite{Sirunyan:2018fhl} Collaborations.
The evidence was obtained from \PbPb data recorded in 2015 at a centre-of-mass energy of $\sqn~=~5.02$~TeV with integrated luminosities of 0.48~$\textrm{nb}^{-1}$~(ATLAS) and  0.39~$\textrm{nb}^{-1}$~(CMS).
The CMS Collaboration also set upper limits on the cross section for ALP production, $\gamma\gamma\rightarrow a \rightarrow \gamma\gamma$, over a mass range of 5--90~\GeV.
Exploiting a data sample of \PbPb\ collisions collected in 2018 at the same centre-of-mass energy with an integrated luminosity of 1.73~$\textrm{nb}^{-1}$, the ATLAS Collaboration observed LbyL scattering with a significance of  $8.2\sigma$~\cite{Aad:2019ock}. These two ATLAS measurements used tight requirements on the diphoton invariant mass~($>6$~\GeV) and single-photon transverse energy~($>3$~\GeV).
 
This paper presents a measurement of the cross sections for $\textrm{Pb+Pb}\,(\gamma\gamma)\rightarrow \textrm{Pb}^{(\ast)}\textrm{+}\textrm{Pb}^{(\ast)}\,\gamma\gamma$ production at $\sqn~=~5.02$~\TeV\ using a combination of Pb+Pb collision data recorded  in 2015 and 2018 by the ATLAS experiment, corresponding to an integrated luminosity of 2.2~$\textrm{nb}^{-1}$. 
This analysis follows the approach proposed in Ref.~\cite{Enterria:2013yra} and the methodology used in the previous measurements~\cite{Aaboud:2017bwk,Aad:2019ock}. However, as a result of improvements in the trigger efficiency and purity of the photon identification, a broader kinematic range in diphoton invariant mass~($>5$~\GeV) and single-photon transverse energy~($>2.5$~\GeV) is covered. This extension results in an increase of about 50\% in expected signal yield in comparison with the previous tighter requirements.
 
The integrated fiducial cross section and four differential distributions involving kinematic variables of the final-state photons are measured.
Two of the distributions characterise the energy of the process: the invariant mass of the diphoton system, $\Minvgg$, and the average transverse momentum of two photons, $(\pt^{\gamma1}+\pt^{\gamma2})/2$. The remaining ones probe angular correlations of the $\gamma\gamma$ system. These are the rapidity\footnote{Rapidity is defined as  $y=\frac{1}{2}\ln{\frac{E+p_z}{E-p_z}}$, where $E$ and $p_z$ are particle's energy and the component of momentum along the beam axis, respectively.} of the diphoton system, $y_{\gamma\gamma}$, and $|\cos(\theta^*)|$, defined as:
\begin{linenomath*}
\begin{equation*}
|\cos(\theta^*)| = \left|\tanh\left(\frac{\Delta y_{\gamma 1,\gamma 2}}{2} \right) \right|,
\end{equation*}
\end{linenomath*}
where $\theta^*$ is the $\gamma\gamma$ scattering angle in the $\gamma\gamma$ centre-of-mass frame, and $\Delta y_{\gamma 1,\gamma 2}$ is the difference between the rapidities of the photons.
 
The measured diphoton invariant mass distribution is used to set  limits on
ALP production via the process $\gamma\gamma\rightarrow a \rightarrow \gamma\gamma$.

\section{\label{sec:Atlas}ATLAS detector}

The ATLAS detector~\cite{ATLAS_detector_paper} at the LHC covers nearly the entire solid angle around the collision point.
It consists of an inner tracking detector surrounded by a thin superconducting solenoid, EM and hadronic calorimeters,
and a muon spectrometer incorporating three large superconducting toroid magnets.
The inner-detector system (ID) is immersed in a \SI{2}{\tesla} axial magnetic field
and provides charged-particle tracking in the pseudorapidity\footnote{\AtlasCoordFootnote} range $|\eta| < 2.5$.
 
The high-granularity silicon pixel detector (Pixel) covers the collision region.
Typically, it provides four measurements per track,
with the first hit being in the insertable B-layer~(IBL)~\cite{ATLAS-TDR-19, Abbott:2018ikt}, which was installed at a mean distance of 3.3~cm from the beam pipe before the start of Run~2.
It is followed by the silicon microstrip tracker (SCT), which usually provides four two-dimensional measurement points per track.
These silicon detectors are complemented by the transition radiation tracker,
which enables radially extended track reconstruction up to $|\eta| = 2.0$.
 
The calorimeter system covers the pseudorapidity range $|\eta| < 4.9$.
Within the region $|\eta|< 3.2$, EM calorimetry is provided by barrel and
endcap lead/liquid-argon (LAr) EM calorimeters (high-granularity for $|\eta| < 2.5$),
with an additional thin LAr presampler covering $|\eta| < 1.8$
to correct for energy loss in material upstream of the calorimeters.
Hadronic calorimetry is provided by the steel/scintillator-tile calorimeter,
segmented into three barrel structures within $|\eta| < 1.7$, and two copper/LAr hadronic endcap calorimeters.
The solid angle coverage is completed with forward copper/LAr and tungsten/LAr calorimeter modules (FCal)
optimised for EM and hadronic measurements respectively.
 
The muon spectrometer (MS) comprises high-precision tracking chambers measuring the deflection of muons in a magnetic field generated by the superconducting air-core toroids.
The precision chamber system covers the region $|\eta| < 2.7$ with three layers of monitored drift tubes,
complemented by cathode strip chambers in the forward region, where the background is highest.
 
The ATLAS minimum-bias trigger scintillators (MBTS) consist of scintillator slats positioned between the ID and the endcap calorimeters, with each side having an outer ring of four slats segmented in azimuthal angle, covering $2.07 < |\eta| < 2.76$, and
an inner ring of eight slats, covering $2.76 < |\eta| < 3.86$.
 
The ATLAS zero-degree calorimeters (ZDC) consist of four longitudinal compartments on each side of the interaction point~(IP), each with one nuclear interaction length of tungsten absorber, with the Cerenkov light read out by 1.5~mm quartz rods.
The detectors are located 140 m from the nominal IP in both directions, covering $|\eta|>8.3$.
 
The ATLAS LUCID-2 detector~\cite{Avoni:2018iuv} consists of 32 photomultiplier tubes
for luminosity measurements and luminosity monitoring.  Its two modules are placed symmetrically at about $\pm17$~m from the nominal IP.
 
The ATLAS trigger system~\cite{Aaboud:2016leb} consists of a Level-1 trigger implemented using a combination of dedicated electronics and programmable logic, and a software-based high-level trigger (HLT).
\newenvironment{DIFnomarkup}{}{}
\begin{DIFnomarkup}
\section{\label{sec:DataMC}Data and Monte Carlo simulation samples}
\end{DIFnomarkup}
 
The data used in this measurement is from \PbPb collisions with a centre-of-mass energy of $\sqn~=~5.02$~\TeV, recorded in 2015 and 2018 with the ATLAS detector at the LHC.
The full data set corresponds to an integrated luminosity of 2.2~\invnb.
Only high-quality data with all detectors operating normally are analysed.
 
Monte Carlo~(MC) simulated events for the LbyL signal process were generated at leading order~(LO) using SuperChic~v3.0~\cite{Harland-Lang:2018iur}.
They take into account box diagrams with leptons and quarks (such as the diagram in Figure~\ref{fig:UPC}), and $W^{\pm}$ bosons, including interference effects. The $W^\pm$ contribution is only important for diphoton masses $m_{\gamma\gamma} > 2 m_{W}$.
Next-to-leading-order QCD and QED corrections are not included. They increase the \gggg\ cross section by a few percent~\cite{Bern:2001dg, Klusek-Gawenda:2016nuo}. 
An alternative LbyL signal sample was generated using calculations from Ref.~\cite{Klusek-Gawenda:2016euz}.
The difference between the nominal and alternative signal prediction is mainly in the implementation of the non-hadronic overlap condition of the Pb ions.
In SuperChic~v3.0 the probability for exclusive $\gamma\gamma$ interactions turns on
smoothly for Pb+Pb impact parameters in the range of 15--20~fm and it is unity for larger values, while the alternative prediction fully suppresses these interactions for impact parameters below 14~fm when two nuclei overlap during the collision. This difference leads to a fiducial cross section for LbyL scattering that is by about 3\% larger in the alternative calculation than in the prediction from SuperChic~v3.0.

The exclusive diphoton final state can also be produced via the strong interaction through a quark loop in the exchange of two gluons in a colour-singlet state.
This central exclusive production~(CEP) background contribution, $gg\rightarrow\gamma\gamma$, was modelled using SuperChic~v3.0.
Background from two-photon production of quark--antiquark pairs was estimated using \textsc{Herwig}++ 2.7.1~\cite{Bahr:2008pv} where the Equivalent Photon Approximation~(EPA) formalism in \pp collisions is implemented. The sample was then normalised to cover the differences in equivalent photon fluxes between the \PbPb\ and $pp$ cases.
 
Exclusive dielectron pairs from the reaction $\textrm{Pb+Pb}\,(\gamma\gamma)\rightarrow \textrm{Pb}^{(\ast)}\textrm{+}\textrm{Pb}^{(\ast)}\,e^+e^-$ are used for various aspects of the analysis, in particular to validate the EM calorimeter energy scale and resolution.
This \yyee\ process was modelled with the STARlight~v2.0 MC generator~\cite{Klein:2016yzr},
in which the cross section is computed by combining the Pb+Pb photon flux with the LO formula for \ggee.
The background contribution from a related process, $\gamma\gamma\rightarrow\tau^+\tau^-$, was modelled using STARlight~v2.0 interfaced with Pythia 8.212~\cite{Sjostrand:2014zea}  for the simulation of $\tau$-lepton decays.
 
Events for the ALP signal were generated using STARlight~v2.0 for ALP masses~($\ma$) ranging between 5 and 100 \GeV. A mass spacing of 1~\GeV\ was used for $5<\ma<30$~\GeV, while for $\ma>30$~\GeV\ a 10~\GeV\ mass spacing was used.
The width of the simulated ALP resonance is well below the detector resolution in all
simulated samples.
 
All generated events were passed through a detector simulation~\cite{Aad:2010ah} based on  GEANT4~\cite{Agostinelli:2002hh} and are reconstructed with the standard ATLAS reconstruction software.

\section{\label{sec:Selection}Event selection}
 
Candidate diphoton events were recorded  using a dedicated trigger for events with moderate activity in the calorimeter but little additional activity in the entire detector. The trigger strategies for the 2015 and 2018 data sets were different. In particular, the latter aimed at improving the trigger efficiency at low photon transverse energy, \et, values.
At Level-1 in 2015, the total \et\ registered in the calorimeter after noise suppression was required to be between 5 and 200~\GeV.
In 2018, a logical OR of two Level-1 conditions was required: (1)~at least one EM cluster with $\et>1\;\gev$ in coincidence with total \et\ registered in the calorimeter between 4--200~\GeV, or (2)~ at least two EM clusters with $\et>1\;\gev$ with total \et\ registered in the calorimeter below 50~\GeV.
At the HLT, events in 2015 were rejected if more than one hit was found in the inner ring of the MBTS (MBTS veto).
In 2018, a requirement of total \et on each side of the FCal detector to be below 3~\GeV\
was imposed.
Additionally, in both data sets a veto condition on activity in the Pixel detector, hereafter referred to as Pixel-veto, had to be satisfied. The number of hits was required to be at most 10 in 2015, and at most 15 in 2018.

Photons are reconstructed from EM clusters in the calorimeter and tracking information provided by the ID, which allows the identification of photon conversions~\cite{Aaboud:2018yqu}.
Selection requirements are applied to remove EM clusters with a large amount of energy from poorly functioning calorimeter cells, and a timing requirement is made to reject out-of-time candidates. An energy calibration specifically optimised for photons~\cite{Aad:2019tso} is applied to the candidates to account for upstream energy loss and both lateral and longitudinal shower leakage. The calibration is derived for nominal $pp$ collisions with dedicated factors applied to account for a negligible contribution from multiple Pb+Pb collisions at the same bunch crossing.
A correction~\cite{Aad:2019tso} is applied to photons in MC samples to account for potential mismodelling of quantities which describe shower shapes of the associated EM showers.
 
The photon particle identification~(photon PID) in this analysis is based on a selection of the shower-shape variables, optimised for the signal events.
Only photons with $\et>2.5~\gev$ and $|\eta|<2.37$, excluding the calorimeter transition region $1.37<|\eta|<1.52$, are considered. The pseudorapidity requirement ensures that the photon candidates pass through regions of the EM calorimeter where the first layer is segmented into narrow strips, providing good separation between genuine prompt photons and photons coming from the decay of neutral hadrons.
The identification is based on a neural network trained on background photons extracted from data and photons from the signal MC simulation, as already used in the previous ATLAS measurement~\cite{Aad:2019ock}. The PID requirements are optimised for low-\et photons~($\et<20$~\GeV) to maintain a constant photon PID efficiency of 95\% as a function of $\eta$ and $\et$ with respect to reconstructed photon candidates. They also select a purer sample of photons than obtained with the cut-based photon PID utilised in \pp\ collisions~\cite{Aaboud:2018yqu}.

Preselected events are required to have exactly two photons satisfying the above selection criteria, with a diphoton invariant mass greater than 5~\GeV.
In order to suppress the \ggee\ background, a veto on charged-particle tracks (with $\pt>100$~\MeV, $|\eta|<2.5$, at least one hit in the Pixel detector and at least six  hits in the Pixel and SCT detectors in total) is imposed.
In order to reduce the background from electrons with poorly reconstructed tracks,
candidate events are required to have no `pixel tracks' in the vicinity of the photon candidate.
Pixel tracks are reconstructed using only the information from the Pixel detector, and are required to have $\pt>50$~\MeV, $|\eta|<2.5$, and at least three hits in the Pixel detector.
In order to suppress fake pixel tracks due to noise in the Pixel detector, only pixel tracks with $\Delta\eta<0.5$ from the photons are considered.
These requirements reduce the fake-photon background from the dielectron final state by a factor of about $10^4$, according to simulation.
They have minor impact on \gggg\ signal events (93\% efficiency for the track veto and 99\% for the pixel-track veto), since the probability of photon conversion in the Pixel detector is relatively small and  the converted photons have a low probability of being reconstructed  at very low \et\ due to the presence of low-momentum electron tracks.
 
Due to the absence of tracks in the LbyL signal events, no primary vertex is reconstructed. The photon direction is estimated using the barycentre of the cluster with respect to the origin of the ATLAS coordinate system.
 
To reduce other sources of fake-photon background (involving mainly calorimeter noise and cosmic-ray muons), the transverse momentum of the diphoton system~(\ptgg) is required to be below 1~\GeV\ for $\Minvgg<12~\gev$ and below 2~\GeV\ for $\Minvgg>12~\gev$.
To reduce real-photon background from CEP $gg\rightarrow\gamma\gamma$ reactions, an additional requirement on the diphoton acoplanarity, $\Aco=(1-|\Delta\phi_{\gamma\gamma}|/{\pi})<0.01$, is used.
The CEP process exhibits a significantly broader acoplanarity distribution than the
$\gamma\gamma\rightarrow\gamma\gamma$ process because gluons recoil against the Pb nucleus, which then dissociates.

To select \ggee candidates, events are required to pass the same trigger as in the diphoton selection.
Each electron is reconstructed from an EM energy cluster in the calorimeter matched
to a track in the ID~\cite{Aaboud:2019ynx}.
The electrons are required to have a transverse energy $\et>2.5~\GeV$ and pseudorapidity $|\eta|<2.47$ with the calorimeter transition region $1.37<|\eta|<1.52$ excluded.
They are also required to meet loose identification criteria based on shower-shape and track-quality variables~\cite{Aaboud:2019ynx}.
The \yyee\ events are selected by requiring exactly two oppositely charged electrons, no further charged-particle tracks coming from the interaction region (with the selection requirements as described above), and dielectron acoplanarity below 0.01.

\section{\label{sec:DetCalib}Detector calibration}
 
\subsection{Trigger efficiency\label{sec:trigger}}
 
The trigger sequence used in the analysis consists of three independent requirements: Level-1, MBTS/FCal veto, and the requirement on low activity in the ID.
 
\begin{figure}[b!]
\begin{center}
\includegraphics[width=0.65\textwidth]{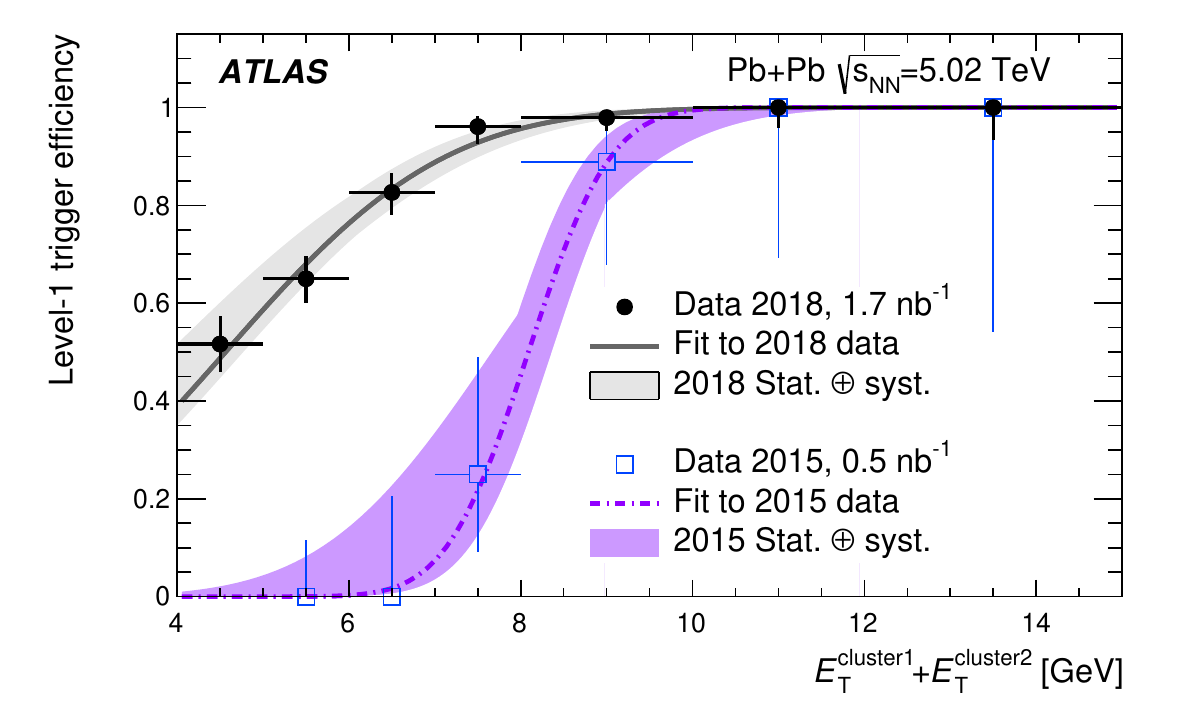}
\caption{ The Level-1 trigger efficiency extracted from \ggee\ events that pass the supporting triggers as a function of the sum of \et of the two EM clusters. Data are shown as points with error bars representing statistical uncertainties, separately for two data-taking periods: 2015 (open squares) and 2018 (full circles). The efficiency is parameterised using the error function fit, shown as a dashed (2015) or solid (2018) line. Shaded bands denote total (statistical and systematic) uncertainty.}
\label{fig:level1_eff}
\end{center}
\end{figure}

The Level-1 trigger efficiency was estimated with \ggee\ events passing one of the independent supporting triggers.
These triggers are designed to select events with single or double dissociation of Pb nuclei and small activity in the ID.
They are based on a coincidence of signals in one or both ZDC sides with a requirement on the total \et\ in the calorimeter to be below 50~\GeV.
Dielectron event candidates are required to have exactly two reconstructed tracks and two geometrically matched EM clusters,  each with a minimum \et\ of 1~\GeV{} and $|\eta| <1.47$, excluding the calorimeter transition region $1.37<|\eta|<1.52$.
The electron identification requirements are removed in order to accept more events in this very low \et\ region, where the efficiencies to reconstruct and identify electrons are low.
Furthermore,
dielectron acoplanarity evaluated using electron charged-particle tracks is required to be below 0.01.
The extracted Level-1 trigger  efficiency is provided as a function of the sum of \et of the two EM clusters reconstructed offline~($\sum\et^{\textrm{clusters}} = \etclone+\etcltwo$).
For $\sum\et^{\textrm{clusters}}=5~\gev$  this efficiency, shown in Figure~\ref{fig:level1_eff},  reaches $60\%$ for 2018 trigger settings, while it is consistent with $0\%$ for 2015 trigger settings due to higher trigger thresholds.
The Level-1 trigger efficiency grows to about $25\%$~($95\%$) for $\sum\et^{\textrm{clusters}}=7.5~\gev$ for 2015~(2018) data.
The efficiency plateau is reached around $\sum\et^{\textrm{clusters}}=10~\gev$ for the 2015  data-taking period and around $\sum\et^{\textrm{clusters}}=9~\gev$
for the 2018 one. The error bars associated with the data points represent statistical uncertainties.
The efficiency is parameterised using an error function fit that is used to reweight the MC simulation.
The statistical uncertainty is estimated by varying the fit parameters by their uncertainty values. The systematic uncertainty is estimated using modified \ggee\ selection criteria.
 
The MBTS and FCal veto efficiencies are estimated using \yyee\ events recorded by supporting triggers.
The MBTS veto efficiency is estimated to be $(98 \pm 2)\%$~\cite{Aaboud:2017bwk}  and the FCal veto efficiency is found to be $(99.1 \pm 0.6)\%$. Both efficiencies are independent of kinematics.
 
Due to low conversion probability of signal photons in the Pixel detector,  inefficiency of the Pixel-veto requirement at the trigger level is found to be negligible for diphoton event candidates.
 
The efficiency for selected \yyee\ events to satisfy the Pixel-veto requirement is evaluated using a dedicated
supporting trigger accepting events with at most 15 tracks at the HLT, out of which at least two had $\pt>1$~\GeV. At Level-1, the same trigger condition was applied as in the diphoton trigger. The FCal veto requirement was also imposed at the HLT.
The Pixel-veto efficiency is parameterised using a second-order polynomial as a function of dielectron rapidity, $y_{ee}$.
The efficiency reaches 80--85\% for dielectron rapidity $|y_{ee}|<1$ and drops to 45--50\% at  $|y_{ee}|\approx 2.5$.
This efficiency correction is applied to the \yyee\  MC simulation.
 
\subsection{Photon reconstruction and identification\label{sec:photon-eff}}

The photon reconstruction efficiency is extracted from data using \ggee events, where one of the electrons emits a hard-bremsstrahlung photon when interacting with the material of the detector.
A tag-and-probe method is performed for events  collected by the diphoton trigger with exactly one identified electron and exactly two reconstructed charged-particle tracks.
The electron is considered a tag if it can be matched to one of the tracks with a $\Delta R < 1.0$ requirement.
The electron  is required to have $\eT^{e}>4$~\GeV\ and the track that is unmatched with the electron (trk2) must have $\pT < 1.5$~\GeV.
The electron--trk2 transverse momentum difference is treated as the transverse energy of the probe, since the additional hard-bremsstrahlung photon is expected to have $\eT^{\gamma}\approx (\eT^e-\pT^\textrm{trk2})$.
The $\pT^\textrm{trk2}<1.5$~\GeV\ requirement ensures a sufficient $\Delta R$ separation between the expected photon and the second electron. A hard-bremsstrahlung photon is expected to be within a distance of $\Delta R = 1.0$ around trk2 direction.
Any additional background contribution to the exclusive \yyee\ reaction is found to be very small in Pb+Pb UPC~\cite{Abbas:2013oua}, and therefore it is considered negligible.

\begin{figure}[htb]
\begin{center}
\includegraphics[width=0.45\textwidth]{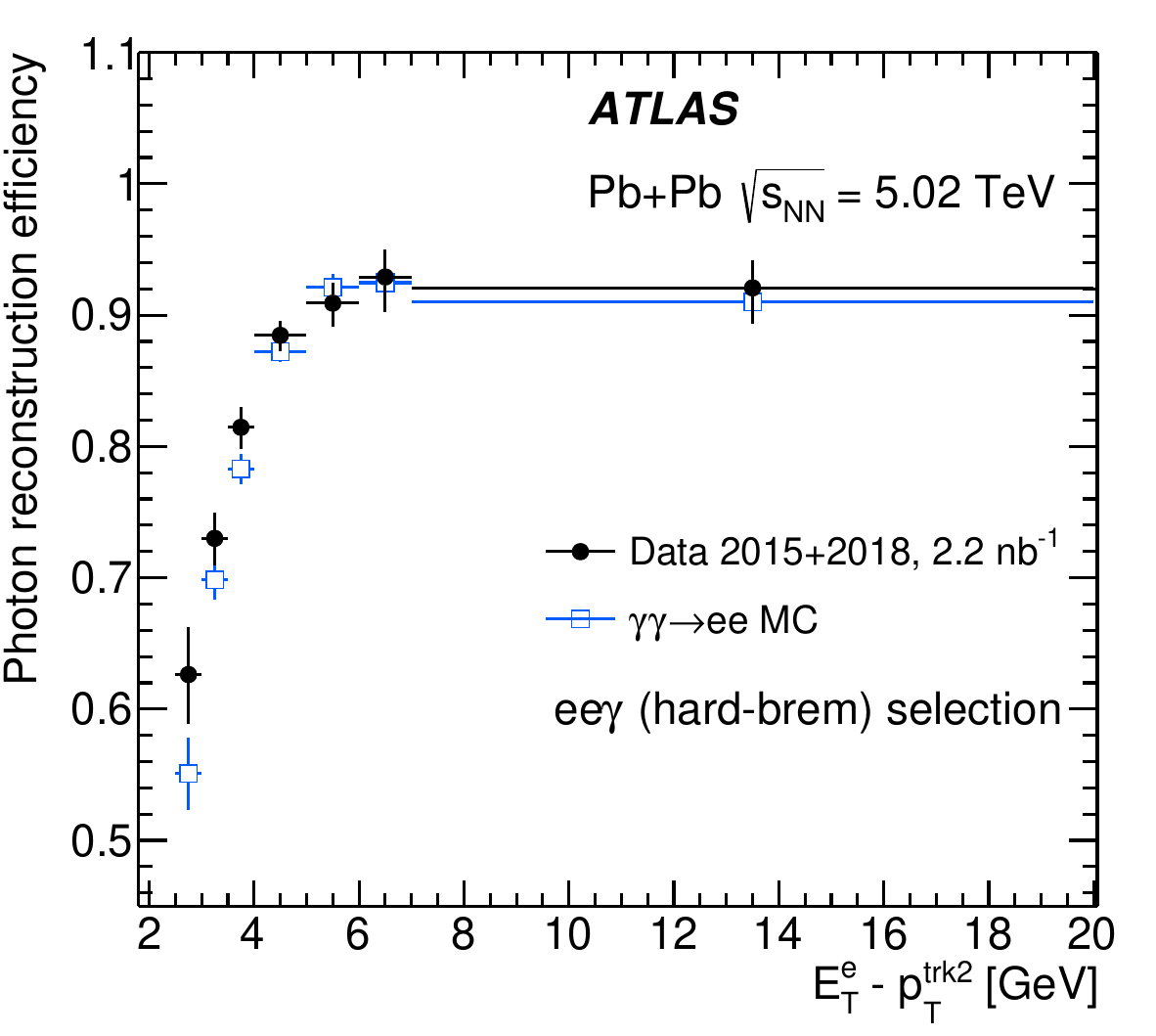}
\caption{Photon reconstruction efficiency as a function of photon $\eT^\gamma$ (approximated with $\eT^e-\pT^\textrm{trk2}$) extracted from \ggee events with a hard-bremsstrahlung photon. Data~(full symbols) are compared with \ggee MC simulation~(open symbols). The error bars denote statistical uncertainties.}
\label{fig:hardbrem_eff2}
\end{center}
\end{figure}

The data sample contains 2905 $\ggee(\gamma)$ bremsstrahlung photons and is used to extract the photon reconstruction efficiency, which is presented in Figure~\ref{fig:hardbrem_eff2}.
The efficiency in data is approximately $60\%$ for $\eT^{\gamma}=2.5~\GeV$ and reaches 90\% at  $\eT^{\gamma}=6~\GeV$.
Reasonable agreement between data and simulation is found.
The distribution from Figure~\ref{fig:hardbrem_eff2} is used to obtain the data-to-simulation scale factors that are used to correct the MC simulation.

High-\pt\ exclusive dilepton production~(\ggll with $\ell^\pm=e^\pm, \mu^\pm$) with final-state radiation~(FSR) is used for data-driven measurements of the photon PID efficiency, defined as the probability for a reconstructed photon to satisfy the identification criteria.
Events with exactly two oppositely charged tracks with $\pt>0.5\;\gev$ are selected in UPC events recorded by the diphoton or dimuon\footnote{ The dimuon trigger  required a muon candidate with $\pt>4$~\GeV\ reconstructed at Level-1 and at least two tracks with \pt above 1~\GeV\ among up to 15 tracks found at the HLT.} triggers. In addition a requirement to reconstruct a photon candidate with $\et^\gamma>2.5$~\GeV\ and $|\eta|<2.37$, excluding the calorimeter transition region $1.37<|\eta|<1.52$, is imposed. A photon candidate is required to be separated from each track with the requirement $\Delta R>0.3$. This condition avoids the leakage of the photon cluster energy to an electron cluster from the \ggee\ process. The mass of the dilepton system is required to be above 1.5~\GeV.
The FSR event candidates are identified using a  $\pTttg<1$~\GeV\ requirement, where \pTttg\ is the transverse momentum of the three body system consisting of two oppositely charged tracks and a photon. The FSR sample consists of 1333~(212) photon candidates in the 2018~(2015) data set and is statistically independent from the hard-bremsstrahlung photon sample used in the photon reconstruction efficiency measurement.
 
\begin{figure}[htb]
\begin{center}
\includegraphics[width=0.45\textwidth]{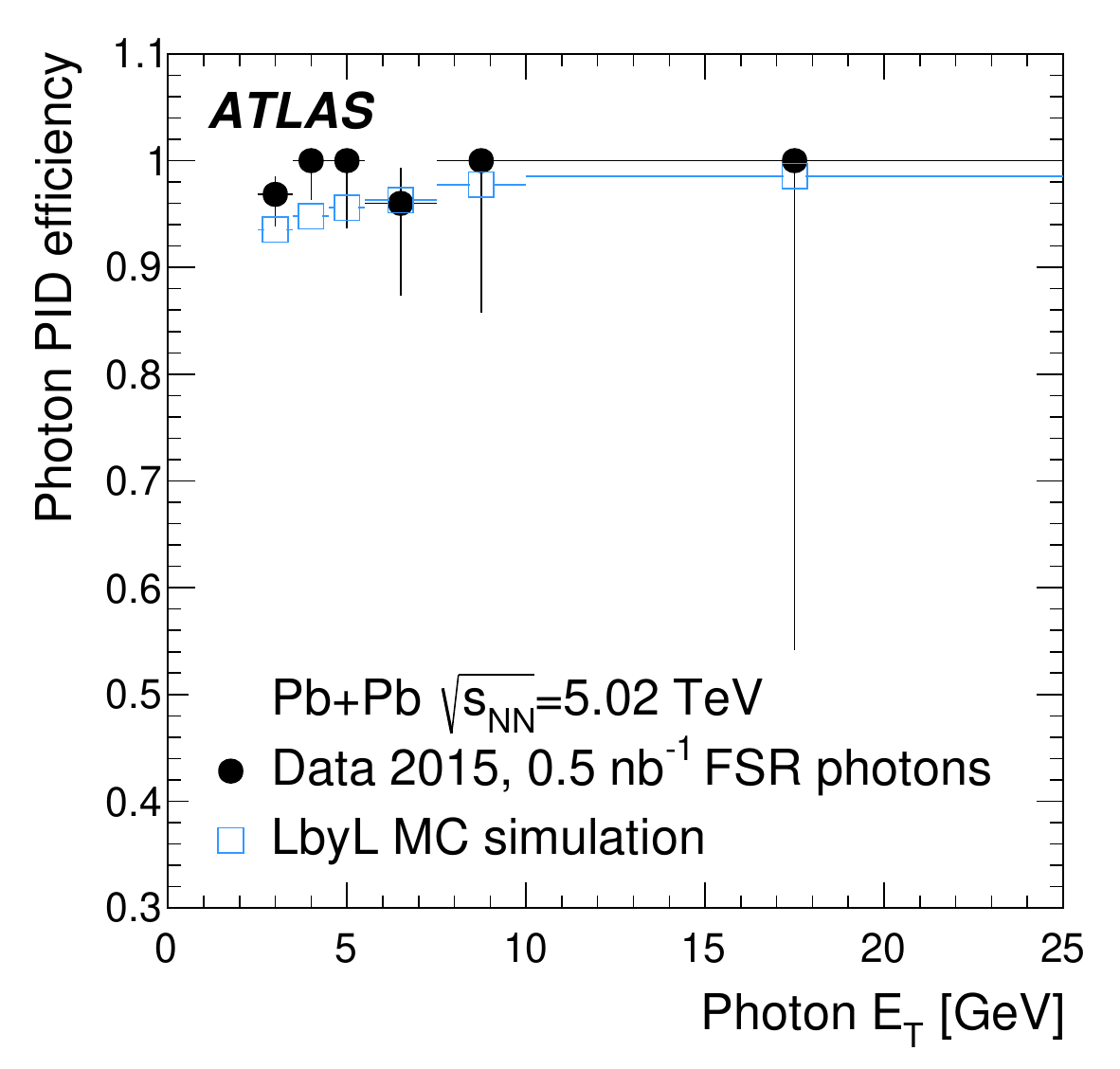}
\includegraphics[width=0.45\textwidth]{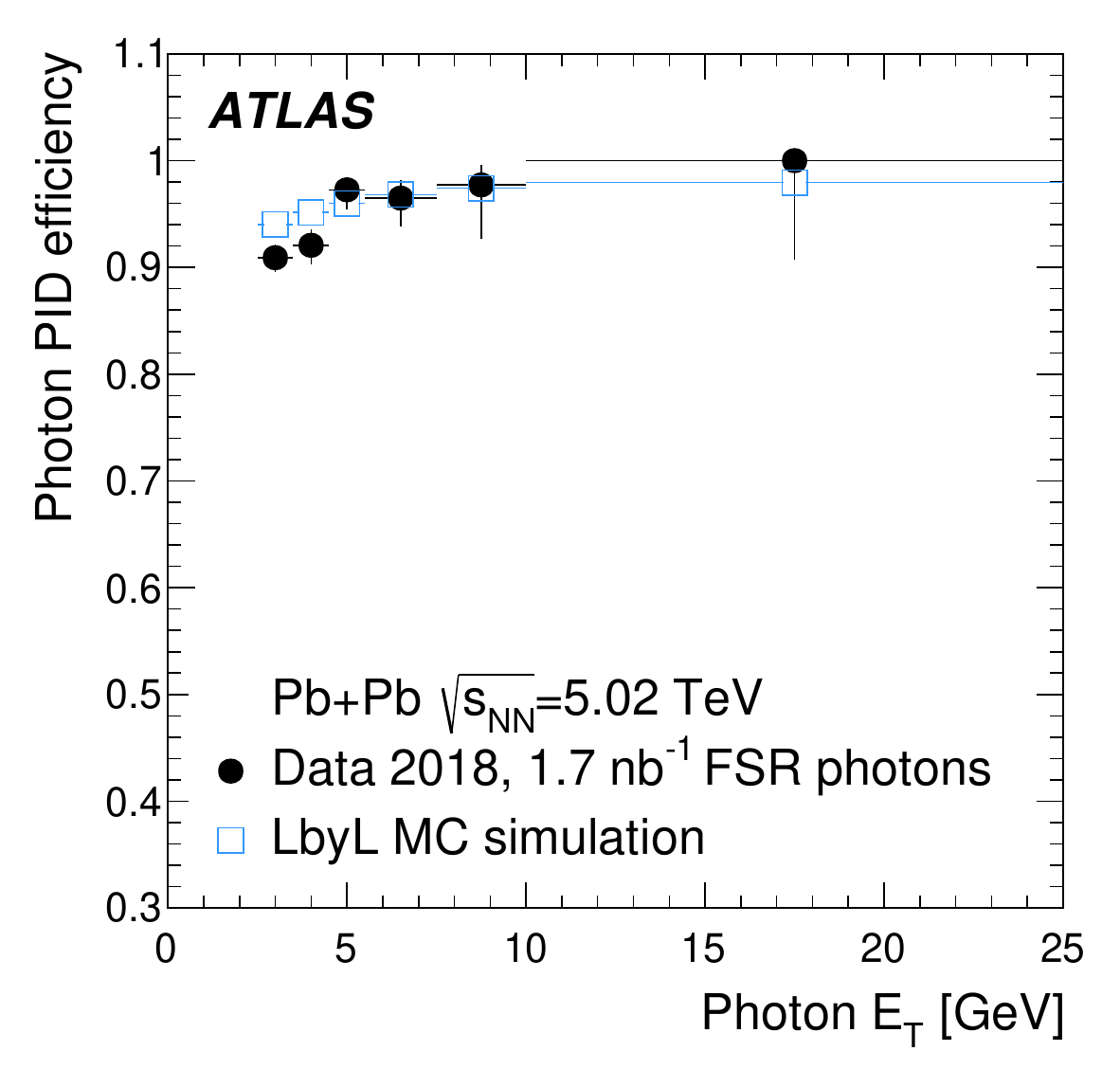}
\caption{Photon PID efficiency as a function of photon \et\ extracted from FSR event candidates in 2015~(left) and 2018~(right) data~(full symbols) and signal MC sample~(open symbols). The error bars denote statistical uncertainties.}
\label{fig:pt-fsr-eff2}
\end{center}
\end{figure}

Figure~\ref{fig:pt-fsr-eff2} shows the photon PID efficiency as a function of the reconstructed photon \et\ for 2015 and 2018 data. The efficiency in data is compared with the efficiency extracted from the signal MC sample.
Photon PID efficiencies in MC simulation with 2015 and 2018 data-taking conditions are in good agreement.
In the data for photons with $\et < 5$~\GeV, the photon PID efficiency is in the range of 91-93\% in the 2018 set, while it is found to be 97-100\% in the 2015 set. This difference is due to slightly different detector conditions between the 2015 and 2018 data-taking periods, causing the photon shower-shape distributions to be narrower in the 2015 data.
Based on these studies, MC simulated events are corrected using photon \et-dependent data-to-simulation scale factors separately for the 2015 and 2018 data sets.

\subsection{Photon energy calibration}
The EM energy scale and energy resolution are validated in data using \ggee\ events.
The two electrons from the \ggee\ reaction exhibit balanced transverse momenta with $|\pt^{e^+}-\pt^{e^-}|$, expected to be below 30~\MeV, which is much smaller than the EM calorimeter energy resolution.
Therefore, the energy resolution, $\sigma_{\et^\textrm{cluster}}$, can be determined from the measurement of $\etclone-\etcltwo$ distributions in \ggee\ events from the formula:
\begin{equation*}
\sigma_{\et^\textrm{cluster}} \approx \frac{\sigma_{(\et^\textrm{cluster1}-\et^\textrm{cluster2})}}{\sqrt{2}}~,
\end{equation*}
where $\etclone$ and $\etcltwo$ are the transverse energies of the two clusters.
At low electron-\et (below 10~\GeV{}) the value of $\sigma_{\et^\textrm{cluster}}/\et^\textrm{cluster}$ is observed to be 8--10\% in data, which agrees well with the resolution obtained from simulation.

The EM energy scale is cross-checked using the ratio of electron cluster \et to electron track \pt.
It is observed that the simulation provides a good description of the $\et^e / \pt^{\textrm{trk}}$ distribution.
 
\subsection{Control distributions for exclusive \yyee\ production }

\begin{figure}[!htb]
\begin{center}
\includegraphics[width=0.45\textwidth]{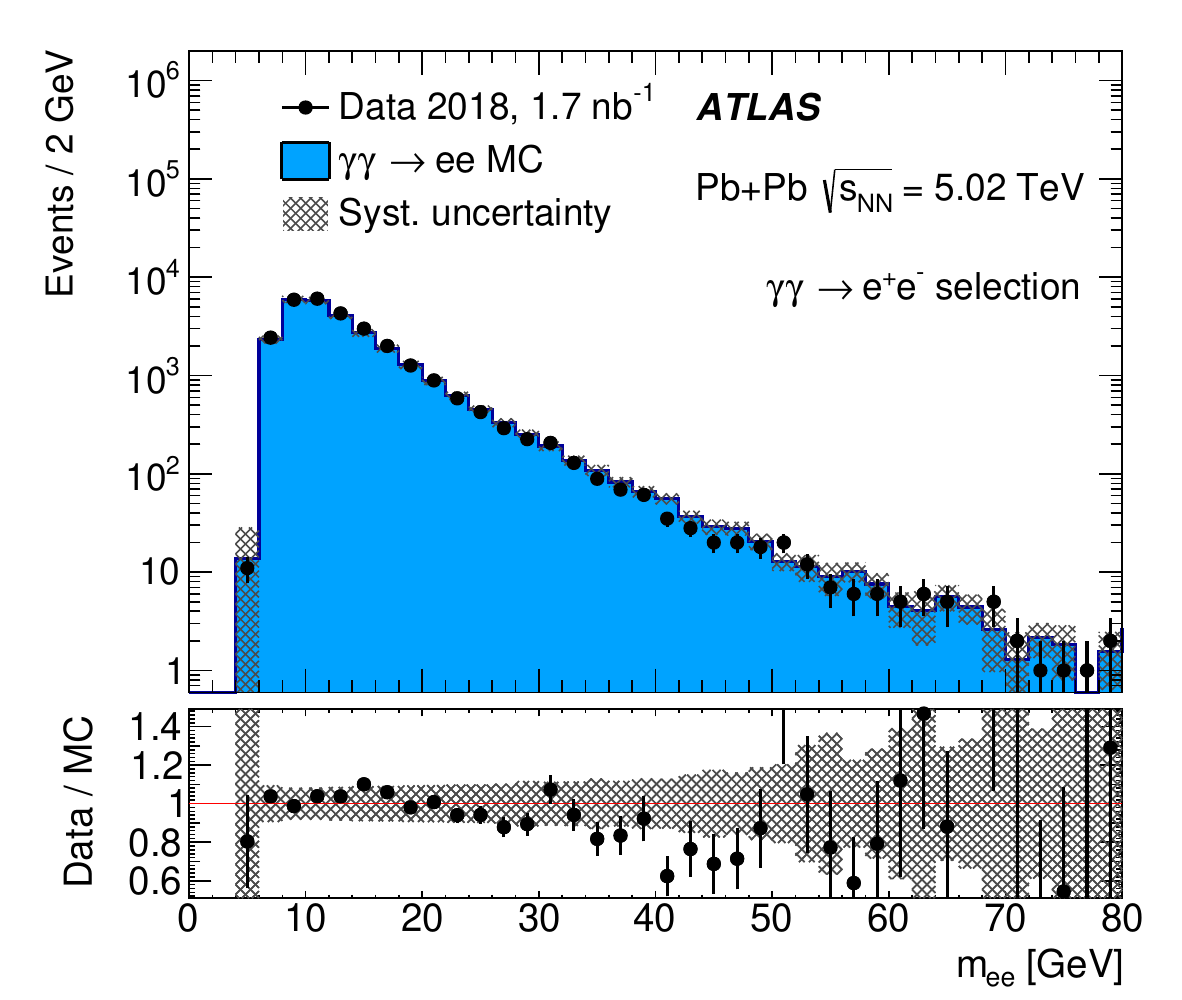}
\includegraphics[width=0.45\textwidth]{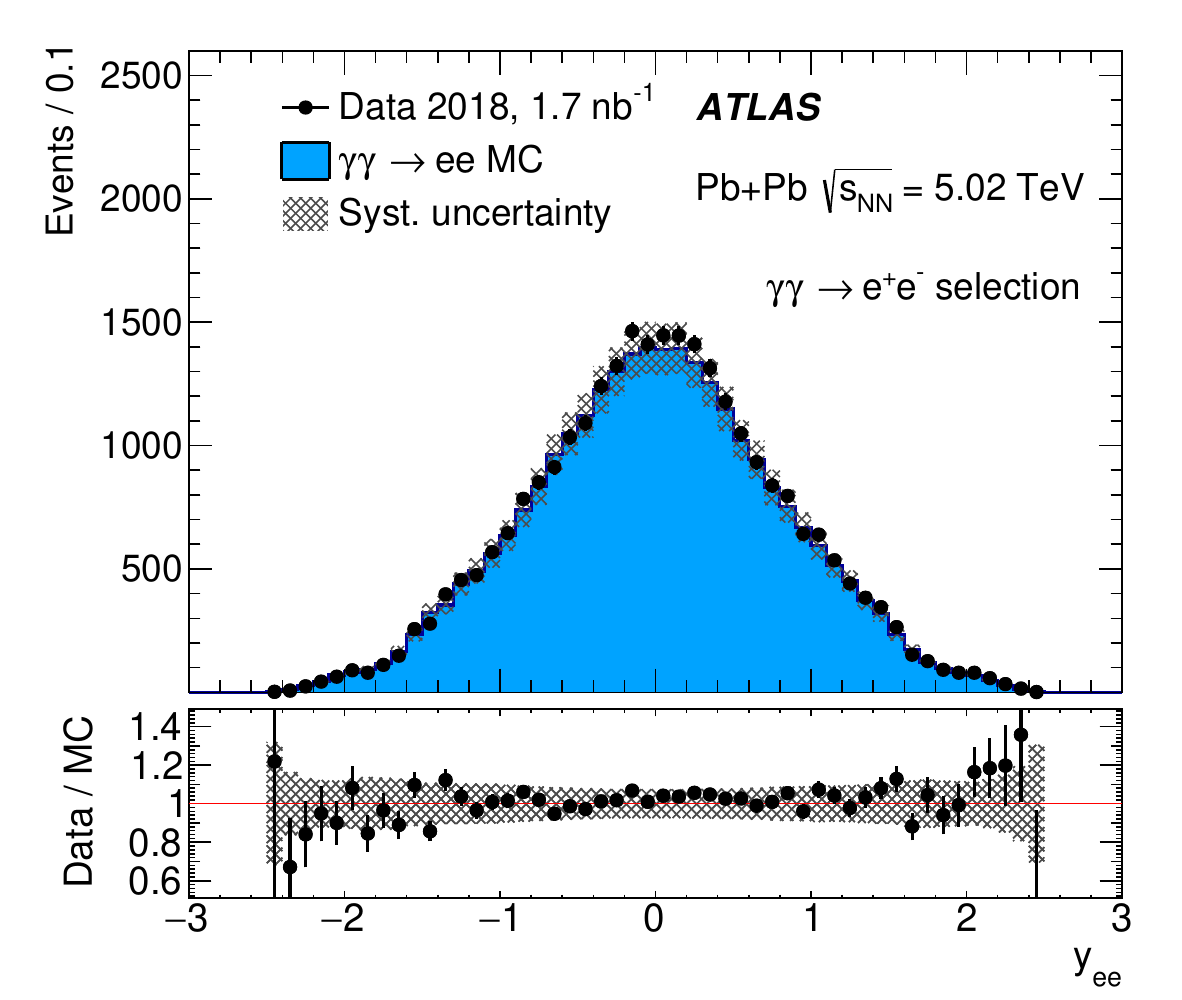}
\includegraphics[width=0.45\textwidth]{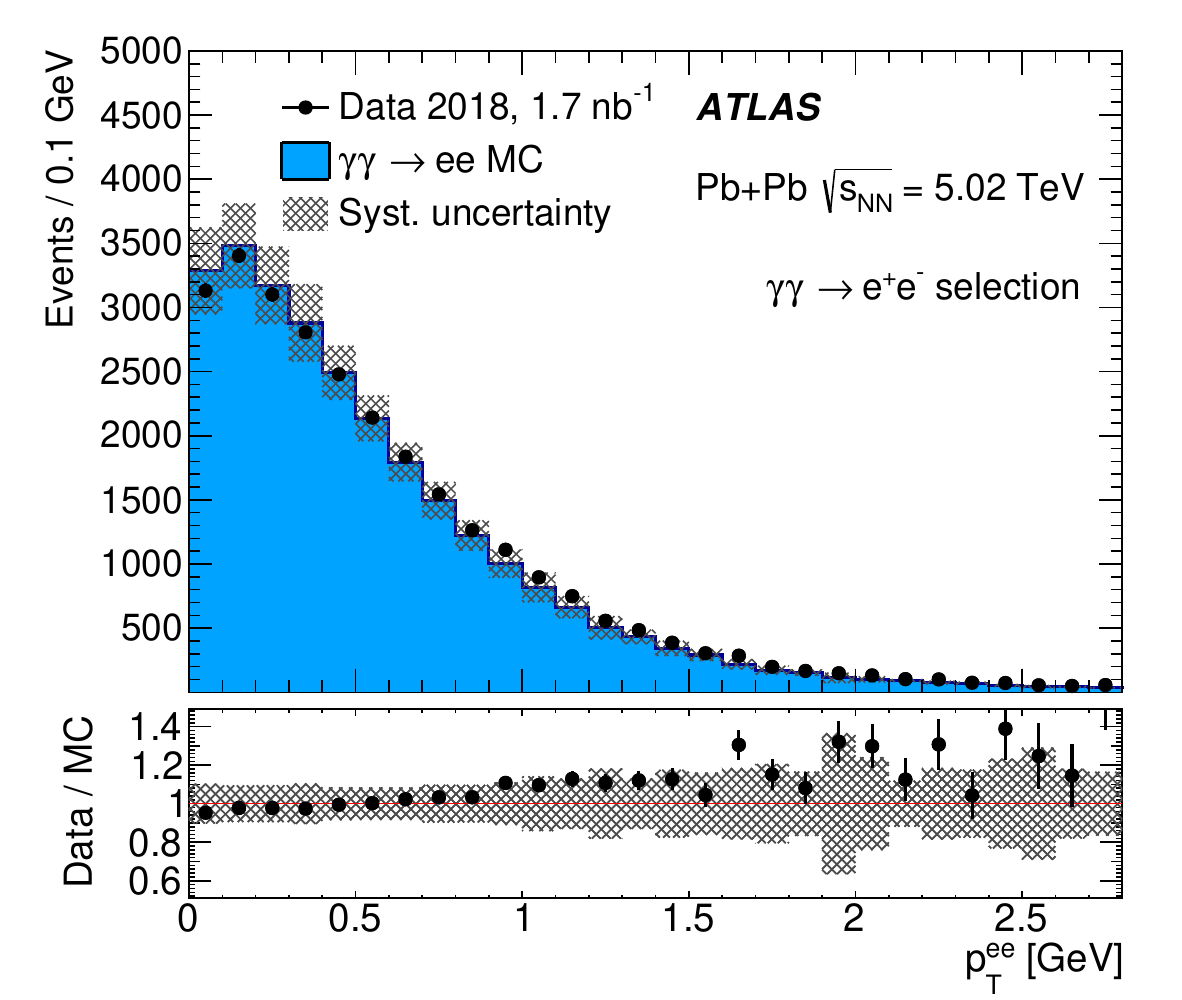}
\includegraphics[width=0.45\textwidth]{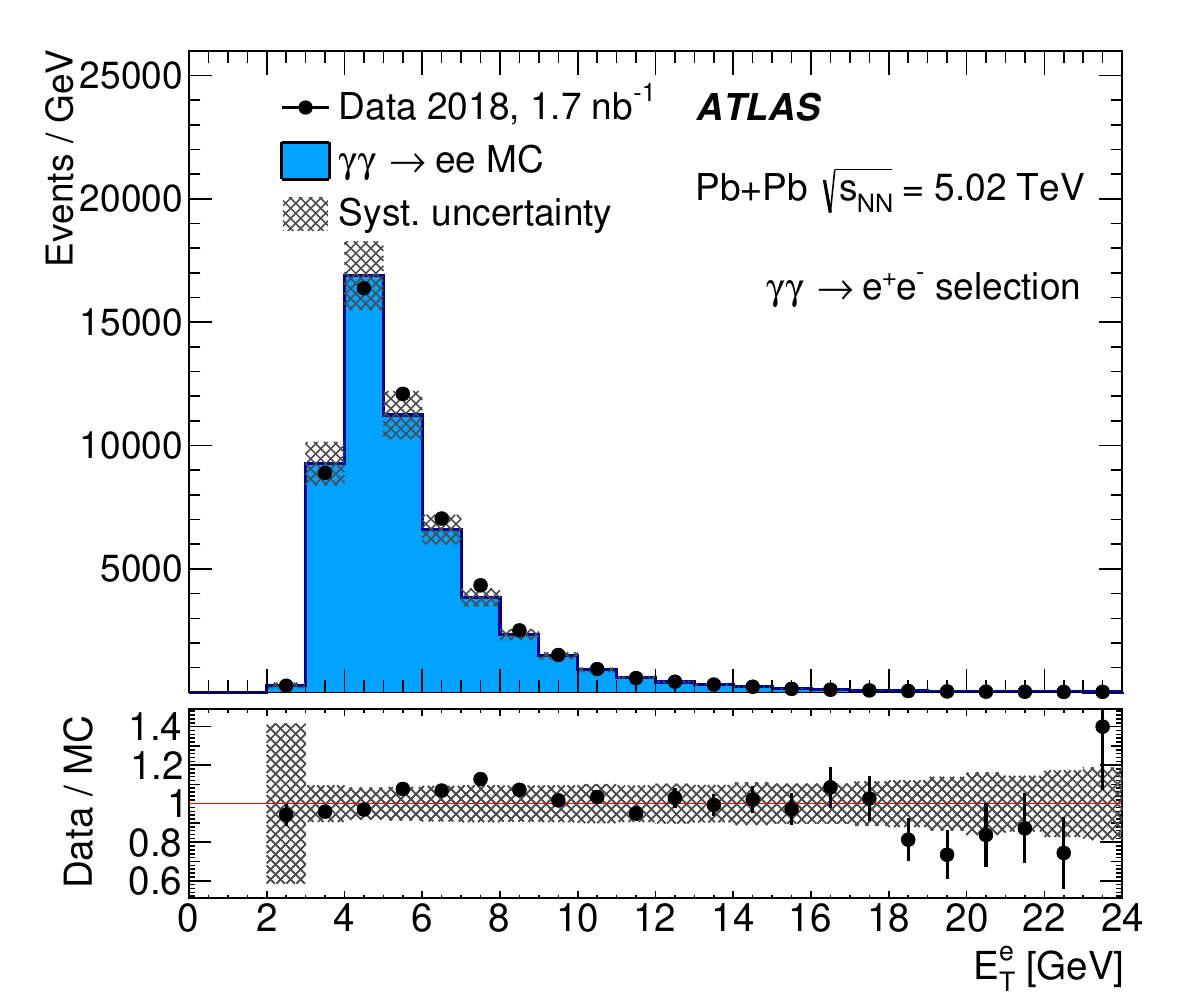}
\caption{Kinematic distributions for $\textrm{Pb+Pb}\,(\gamma\gamma)\rightarrow \textrm{Pb}^{(\ast)}\textrm{+}\textrm{Pb}^{(\ast)}
\,e^+e^-$ event candidates in the 2018 data set: dielectron mass (top-left), dielectron rapidity (top-right), dielectron \pt~(bottom-left) and electron transverse energy (bottom-right).
Data (points) are compared with MC expectations (histograms). The simulation prediction is normalised to the same integrated luminosity as the data.
Systematic uncertainties due to electron energy scale and resolution, electron reconstruction and identification, and trigger efficiency, are shown as shaded bands.
The lower panels display the ratio of data to MC predictions. Some values are outside the plotting range.}
\label{fig:ele-kin_2018}
\end{center}
\end{figure}

Figure~\ref{fig:ele-kin_2018} presents detector-level distributions for events passing the \yyee\ selection (outlined in Section~\ref{sec:Selection}) in the 2018 \PbPb\ data.
In total, 28\,045 \yyee\ event candidates are observed. The shaded bands reflect systematic uncertainties due to electron energy scale and resolution, electron reconstruction and identification, and trigger efficiency.
In general, the STARlight prediction describes the normalisation and shapes of distributions well.
Small systematic differences between the central values of the exclusive dielectron data and the MC prediction are seen in the tail of the dielectron~\pt\ distribution, likely due to a missing contribution from the QED final-state radiation which is not simulated by the MC generator.

The low number of \yyee\ events collected by a control trigger in the 2015 \PbPb\ data precludes precision comparisons between data and MC simulation in that sample.
In particular, the tighter Pixel-veto requirement imposed at the HLT
necessitates a dedicated pseudorapidity-dependent trigger efficiency correction which, due to the limited number of \yyee\ events, could only be extracted with 20\%
precision. Nevertheless, overall reasonable agreement was found within large uncertainties as demonstrated in the previous ATLAS publication~\cite{Aaboud:2017bwk}.

\section{\label{sec:Background}Background estimation}

\subsection{Dielectron final states\label{sec:eeEstimate}}
 
The $\ggee$ process has a relatively high cross section and  can be a source of fake diphoton events.
The electron-to-photon misidentification can occur when the electron track is not reconstructed or the electron emits a hard bremsstrahlung photon.

The $\ggee$ yield in the signal region defined in Section~\ref{sec:Selection} is estimated using a fully data-driven method.
A control region is defined requiring exactly two photon candidates passing the signal selection, and one or two pixel tracks. This control region is denoted by \CROneTwo.
The event yield observed in \CROneTwo\ is extrapolated to the signal region using the probability of missing the electron pixel track if the standard track is not reconstructed (\pmis).
 
The  \pmis value is measured in data
using events with exactly one standard track and two photon candidates having $\Aco < 0.01$.
It is measured to be $\pmis = (47\pm9)\%$, where the uncertainty is estimated by relaxing the \Aco requirement.
It is also found that $\pmis$ does not depend on the probed photon $\et$ and $\eta$.

The number of \ggee events in the signal region is estimated to be $N_{\gamma\gamma\rightarrow e^+e^-} = 15\pm7$, where the uncertainty accounts for the \pmis uncertainty and limited event yield in \CROneTwo.
This uncertainty also covers the differences if the $\ggee$ yield is instead extrapolated from event yields for individual pixel-track multiplicities ($N =1$ or $N =2$).
 
The distribution shapes of various kinematic variables of $\ggee$ background in the signal region are taken from data in \CROne.
The shape uncertainty is constructed by comparing kinematic distributions from data in \CROne\ with the distributions from data in \CRTwo.

\subsection{Central exclusive diphoton production}
 
\begin{figure}[htb]
\begin{center}
\includegraphics[width=0.47\textwidth]{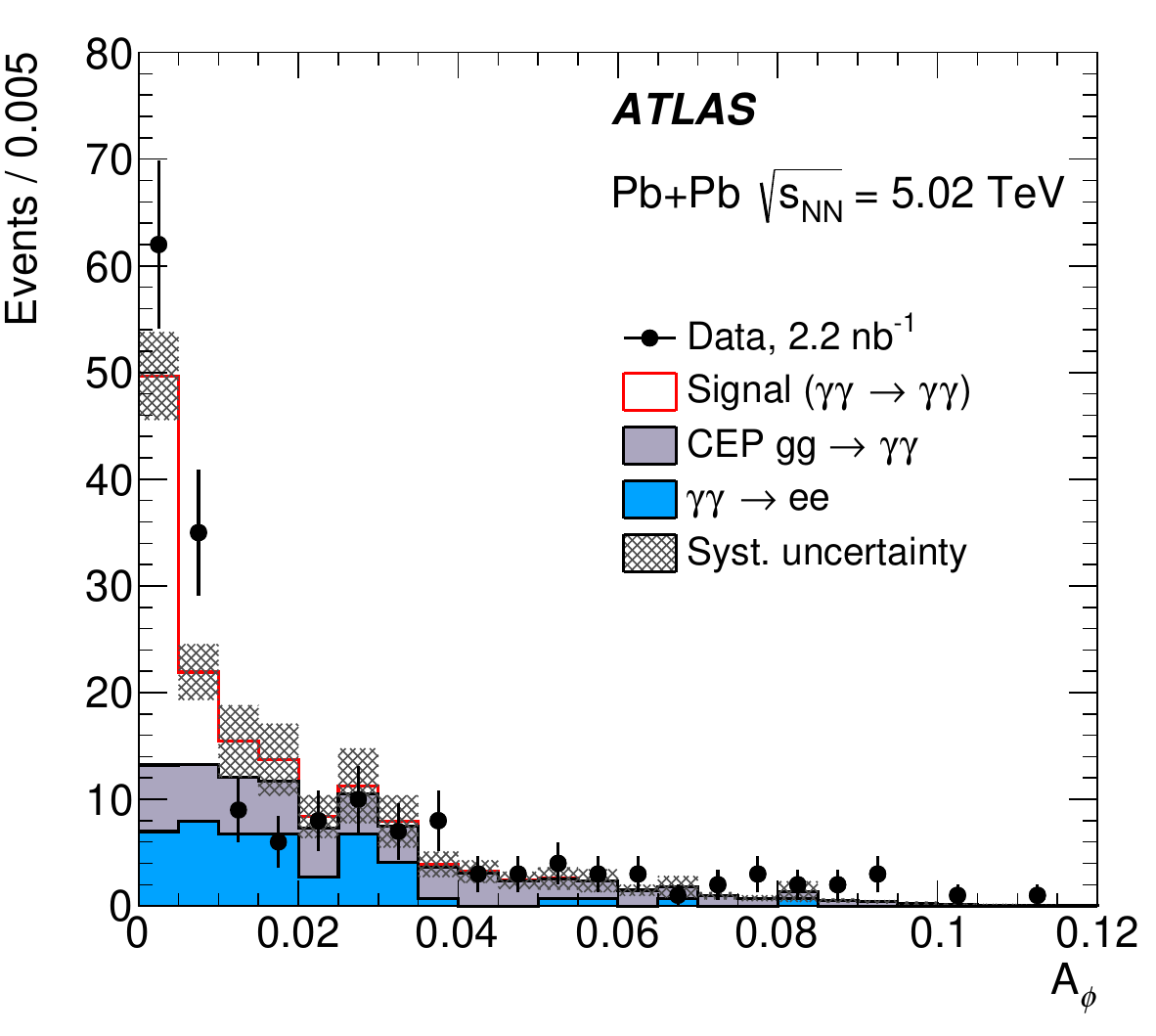}
\caption{The diphoton acoplanarity distribution for events satisfying the signal region selection, but before applying the $\Aco < 0.01$ requirement.
Data are shown as points with statistical error bars, while the histograms represent the expected signal and background levels.
The CEP $gg\rightarrow\gamma\gamma$ background is normalised in the $\Aco > 0.01$ control region.
The signal prediction is normalised to the same integrated luminosity as the data.  The shaded band represents the uncertainties in signal and background predictions, excluding the uncertainty in the luminosity.
}
 
\label{fig:acoNorm}
\end{center}
\end{figure}

The CEP $gg\rightarrow\gamma\gamma$ background is estimated from MC simulation with the overall rate of this process evaluated in the \Aco control region in the data.
The normalisation is constrained using the condition:
\begin{equation*}
N_\textrm{data}(\Aco > 0.01) = N_{gg\rightarrow\gamma\gamma}(\Aco > 0.01) + N_\textrm{sig}(\Aco > 0.01) + N_{\gamma\gamma\rightarrow ee}(\Aco > 0.01)~,
\end{equation*}
where $N_\textrm{data}$  denotes the number of observed events, $N_{gg\rightarrow\gamma\gamma}$ is the expected  CEP $gg\rightarrow\gamma\gamma$ event yield, $N_\textrm{sig}$ is the expected number of signal events  (from MC simulation) and $N_{\gamma\gamma\rightarrow ee}$ is the $e^+e^-$ background yield.
The $N_{\gamma\gamma\rightarrow ee}$ is estimated using the same data-driven method as described in Section~\ref{sec:eeEstimate}.
The diphoton acoplanarity distribution for events satisfying the signal region selection, but before applying the $\Aco < 0.01$ requirement is shown in Figure~\ref{fig:acoNorm}.
The predictions provide a fair description of the shape of the data distribution.
 
The uncertainty in the CEP $gg\rightarrow\gamma\gamma$ background process takes into account the limited number of events in the $\Aco > 0.01$ control region ($11\%$), as well as experimental and modelling uncertainties.
It is found that all experimental uncertainties have a negligible impact on the CEP $gg\rightarrow\gamma\gamma$ background estimate.
The impact of the MC modelling uncertainty on the shape of the acoplanarity distribution is estimated using an alternative SuperChic~v2.0 MC sample with extra gluon interactions (no absorptive effects). This leads to a $21\%$ change in the CEP background yield in the signal region, which is taken as a systematic uncertainty.
An additional check is performed by varying the parton distribution function (PDF) of the gluon. The differences between leading-order MMHT 2014~\cite{Harland-Lang:2014zoa}, CT14~\cite{Dulat:2015mca} and NNPDF3.1~\cite{Ball:2017nwa} PDF sets have negligible impact on the shape of the diphoton acoplanarity distribution.

In addition, the energy deposition in the ZDC, which is sensitive to the dissociation of Pb nuclei,  is studied for events before the $\Aco <0.01$ requirement is imposed.
Good agreement is observed in the $\Aco > 0.01$ control region between the data-driven CEP estimate and the observed events with a signal corresponding to at least one neutron in the ZDC.
In the signal region ($\Aco < 0.01$), approximately 70\% of observed events have a signal corresponding to no neutrons in the ZDC, which is consistent with the signal-plus-background hypothesis.
 
The background due to CEP in the signal region is estimated to be $12\pm3$  events.
In the differential cross-section measurements, the shape uncertainty is evaluated using the alternative SuperChic~v2.0 MC sample.

\begin{DIFnomarkup}
\subsection{Other background sources with prompt photons}
\end{DIFnomarkup}

The contribution from the $\gamma\gamma\rightarrow e^+e^-\gamma\gamma$ process is evaluated using the \textsc{MadGraph5}$\_$aMC$@$NLO v2.4.3 MC generator~\cite{Alwall:2014hca} and the Pb+Pb photon flux from STARlight. This contribution is estimated to be below 1\% of the expected signal and is consequently ignored in the analysis.
 
The contribution from bottomonia production (for example, $\gamma\gamma\rightarrow \eta_b \rightarrow \gamma\gamma$ or $\gamma\textrm{Pb}\rightarrow \Upsilon \rightarrow \gamma\eta_b \rightarrow 3\gamma$) is calculated using relevant branching fractions from Refs.~\cite{Ebert:2002pp, Segovia:2016xqb} and found to be negligible.
 
The contribution from UPC events where both nuclei emit a bremsstrahlung photon is estimated using calculations from Ref.~\cite{Bertulani:1987tz}. The cross section for single-bremsstrahlung photon production from a Pb ion in the fiducial region of the measurement is calculated to be below \SI{e-4}{\pb}, so the coincidence of two such occurrences is negligible.

\subsection{Other fake-photon background}
 
The background contribution from \ggqq\ production is estimated using MC simulation based on \textsc{Herwig}++ and it contributes less than 1 event to the total number of events in the signal region.
The expected yield for the background from $\gamma\gamma\rightarrow\tau^+\tau^-$ process is estimated using MC simulation based on STARlight + Pythia 8 and is found to be less than 0.5 events.
Both of these background sources are considered negligible.
 
Exclusive two-meson production can be a potential source of background for LbyL scattering events, mainly due to their similar back-to-back topology. Mesons can fake photons either by their decay into photons ($\pi^0$, $\eta$, $\eta'$) or by mis-reconstructed charged-particle tracks (for example $\pi^+, \pi^-$ states).
Estimates for such contributions are reported in Refs. \cite{KlusekGawenda:2011ib, Klusek-Gawenda:2019ijn,Enterria:2013yra, HarlandLang:2011qd, Harland-Lang:2013ncy} and these contributions are considered to be negligible in the signal region.

The background from fake diphoton events induced by cosmic-ray muons is estimated using a control region with at least one track reconstructed in the muon spectrometer and further studied using the reconstructed photon-cluster time distribution.
The latter method is also used to estimate the background originating from calorimeter noise.
After imposing the $\pt^{\gamma\gamma}$ requirements, these background contributions are below 1 event and are considered negligible.

\section{\label{sec:Systematics}Systematic uncertainties}

Systematic uncertainties in the $\gamma\gamma\rightarrow\gamma\gamma$ cross-section measurements arise from the reconstruction of photons, the background determination, and integrated luminosity uncertainty, as well as the procedures used to correct for detector effects.
 
The precision of the Level-1 trigger efficiency estimation is limited by the number of events recorded by the supporting trigger.
As a systematic check, the $e^+e^-$ event selection is varied.
In total, the impact of the Level-1 trigger efficiency  uncertainty on the expected signal yield is 5\%.
The uncertainty in the MBTS/FCal veto efficiency has negligible impact on the results.
 
The uncertainty in the photon reconstruction and PID efficiencies is estimated by parameterising the scale factors as a function of the photon pseudorapidity, instead of the photon transverse momentum.
This affects the expected signal yield by 4\% (photon reconstruction efficiency) and 2\% (photon PID efficiency).
The variation of the selection criteria used in data-driven efficiency measurements has negligible impact on the results.
The statistical uncertainty of the photon reconstruction and PID efficiency corrections is propagated using the pseudo-experiment method in which the correction factors are randomly shifted in an ensemble of pseudo-experiments according to the mean and standard deviation of the correction factor.
This has negligible impact on the expected signal.

The uncertainties related to the photon energy scale and resolution  affect the expected signal yield by 1\% and 2\%, respectively.
The uncertainty due to imperfect knowledge of the photon angular resolution is estimated using electron clusters from the $\ggee$ process. The data--MC difference in the electron cluster $\phi$ resolution is applied as an extra smearing to photons from the signal MC sample. This results in a 2\% shift of the signal yield, which is taken as a systematic uncertainty.

The uncertainty due to the choice of signal MC generator is estimated by using an alternative signal MC sample, as detailed in Section~\ref{sec:DataMC}. This affects the signal yield by 1\% which is taken as a systematic uncertainty.
The uncertainty due to the limited signal MC sample size is 1\%.
 
The uncertainties in the background estimation are evaluated as described in Section~\ref{sec:Background}.
 
The uncertainty in the integrated luminosity of the data sample is 3.2\%.
It is derived from the calibration of the luminosity scale using $x$--$y$ beam-separation scans, following a methodology similar to that detailed in Ref.~\cite{ATLAS-CONF-2019-021}, and using the LUCID-2 detector for the baseline luminosity measurements.
 
Systematic uncertainties associated with the background estimate, the photon PID and reconstruction efficiency, photon energy scale, and photon angular and energy resolution are fully correlated between the 2015 and 2018 data-taking periods. Systematic uncertainties in the trigger efficiency are computed separately for each data-taking period. They are dominated by the statistical uncertainty of each data set and are thus uncorrelated.
\section{\label{sec:Results}Results}

\subsection{Kinematic distributions}
 
\begin{figure}[!htb]
\begin{center}
\includegraphics[width=0.42\textwidth]{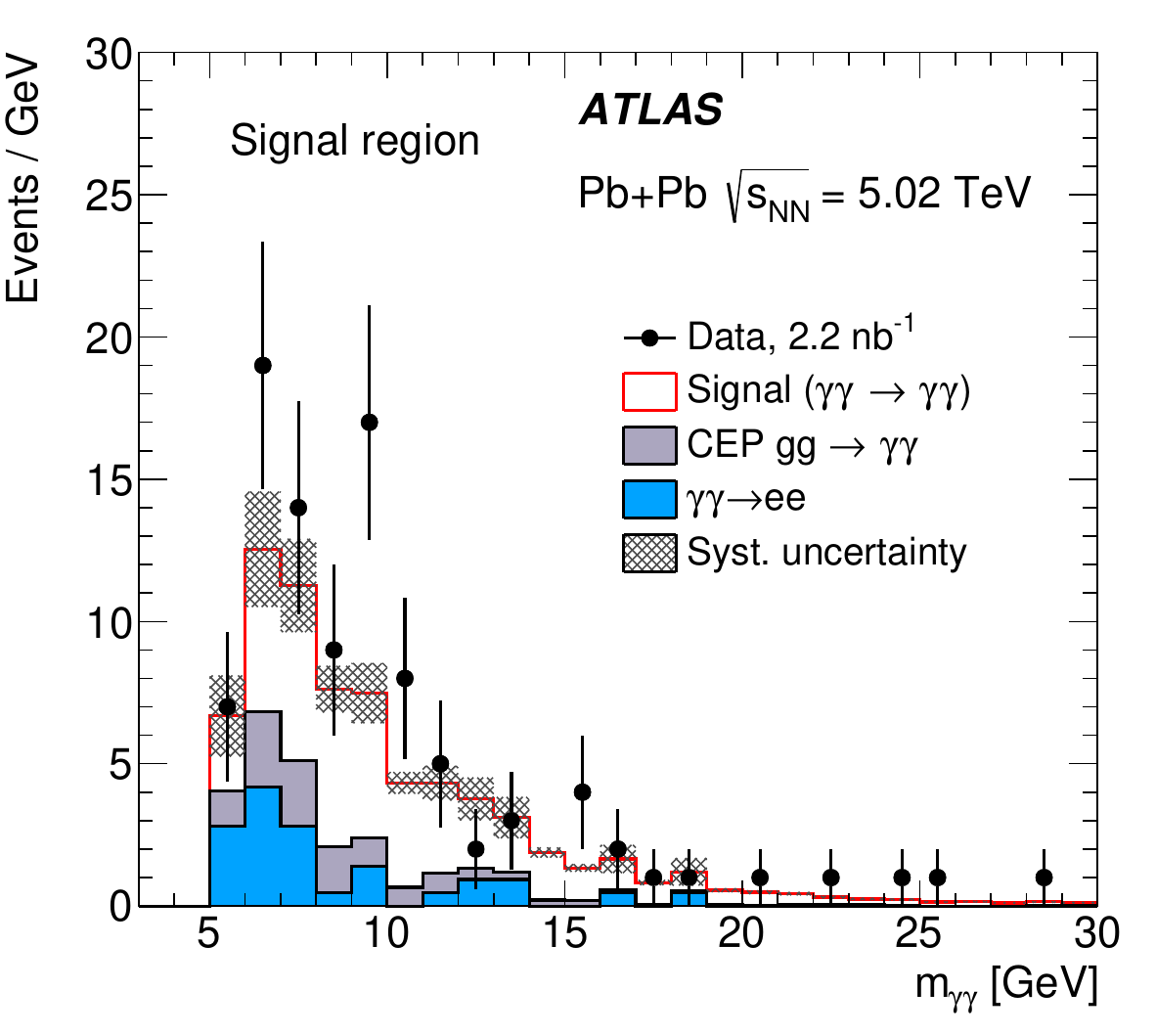}
\includegraphics[width=0.42\textwidth]{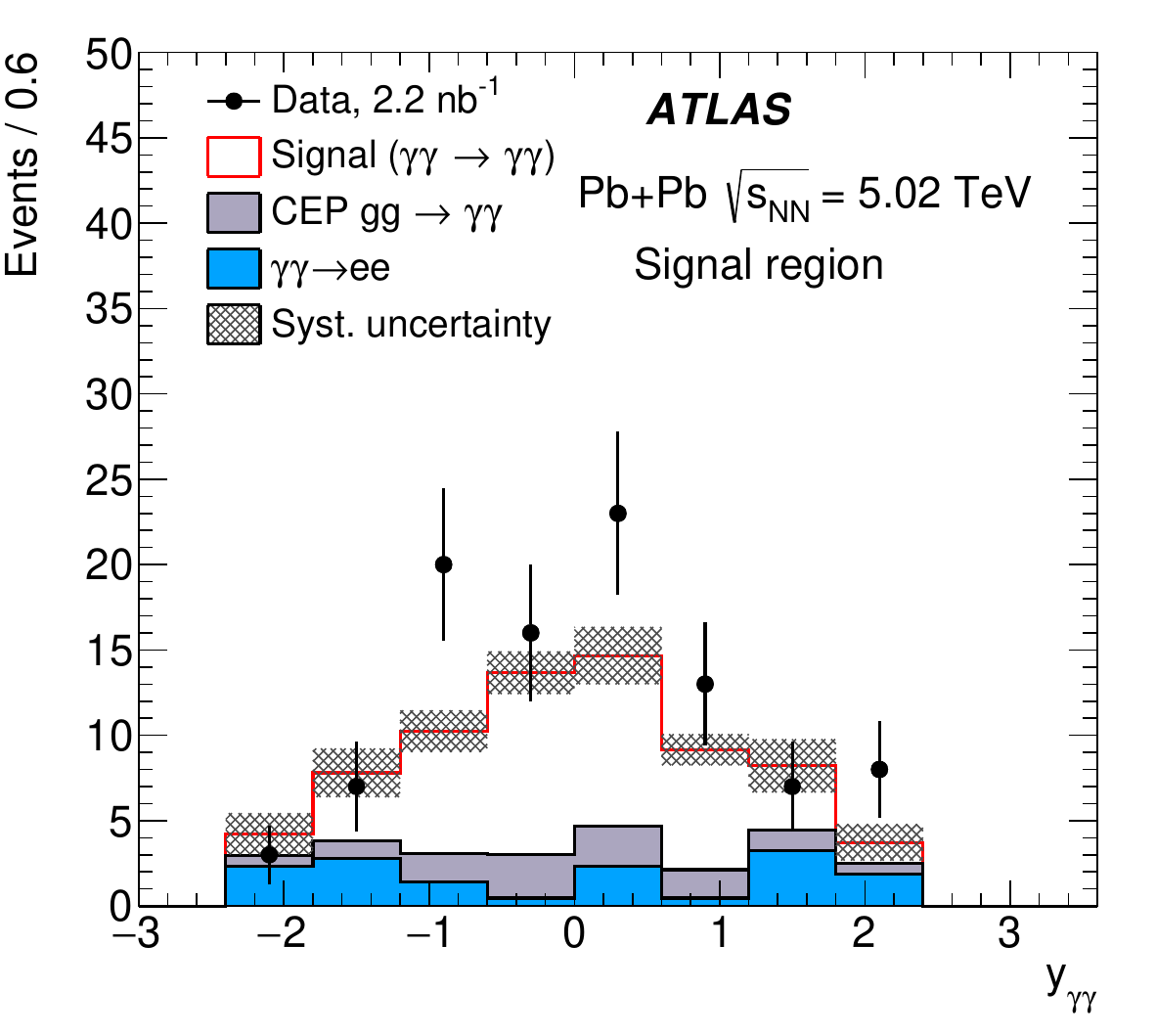}
\includegraphics[width=0.42\textwidth]{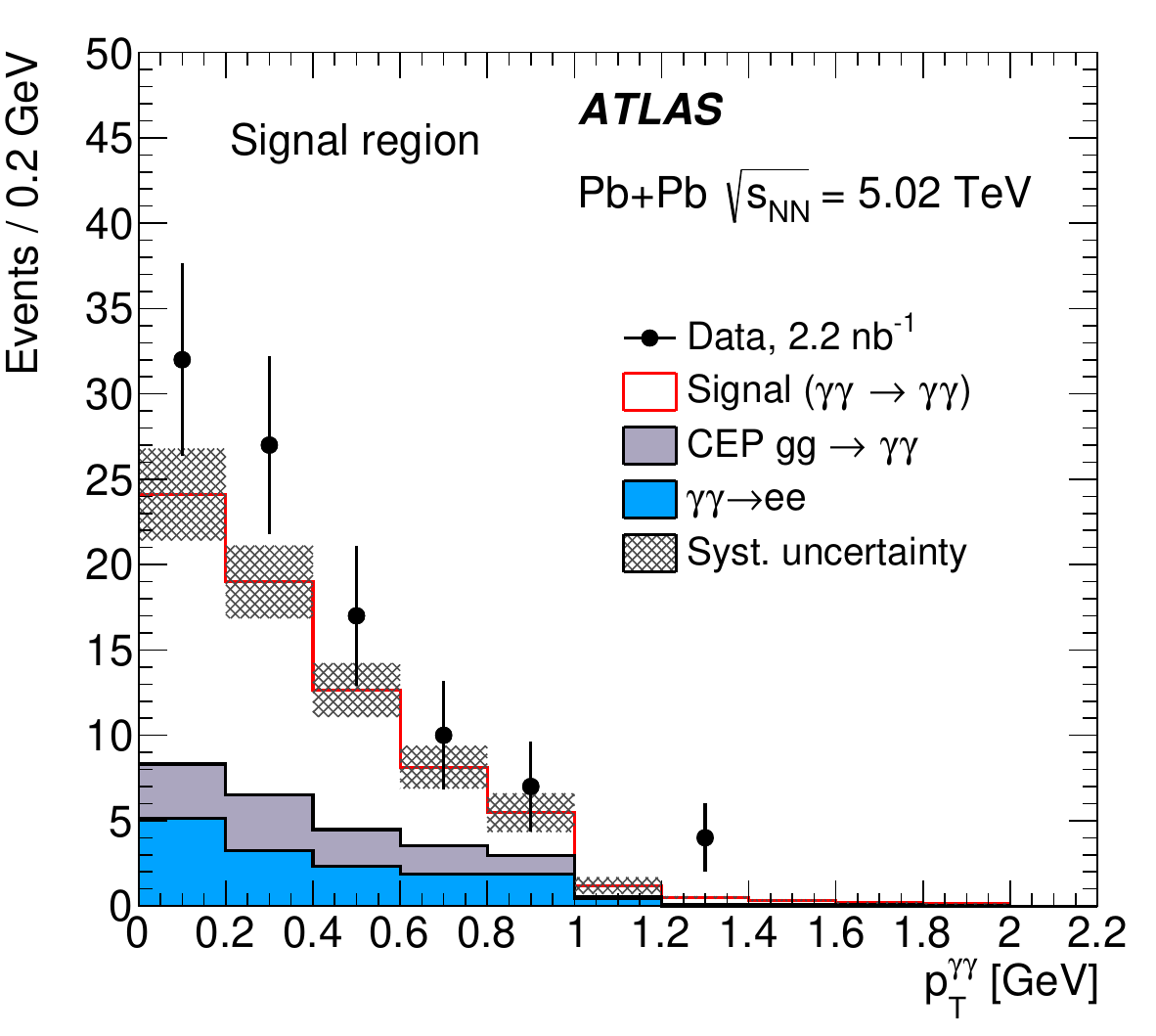}
\includegraphics[width=0.42\textwidth]{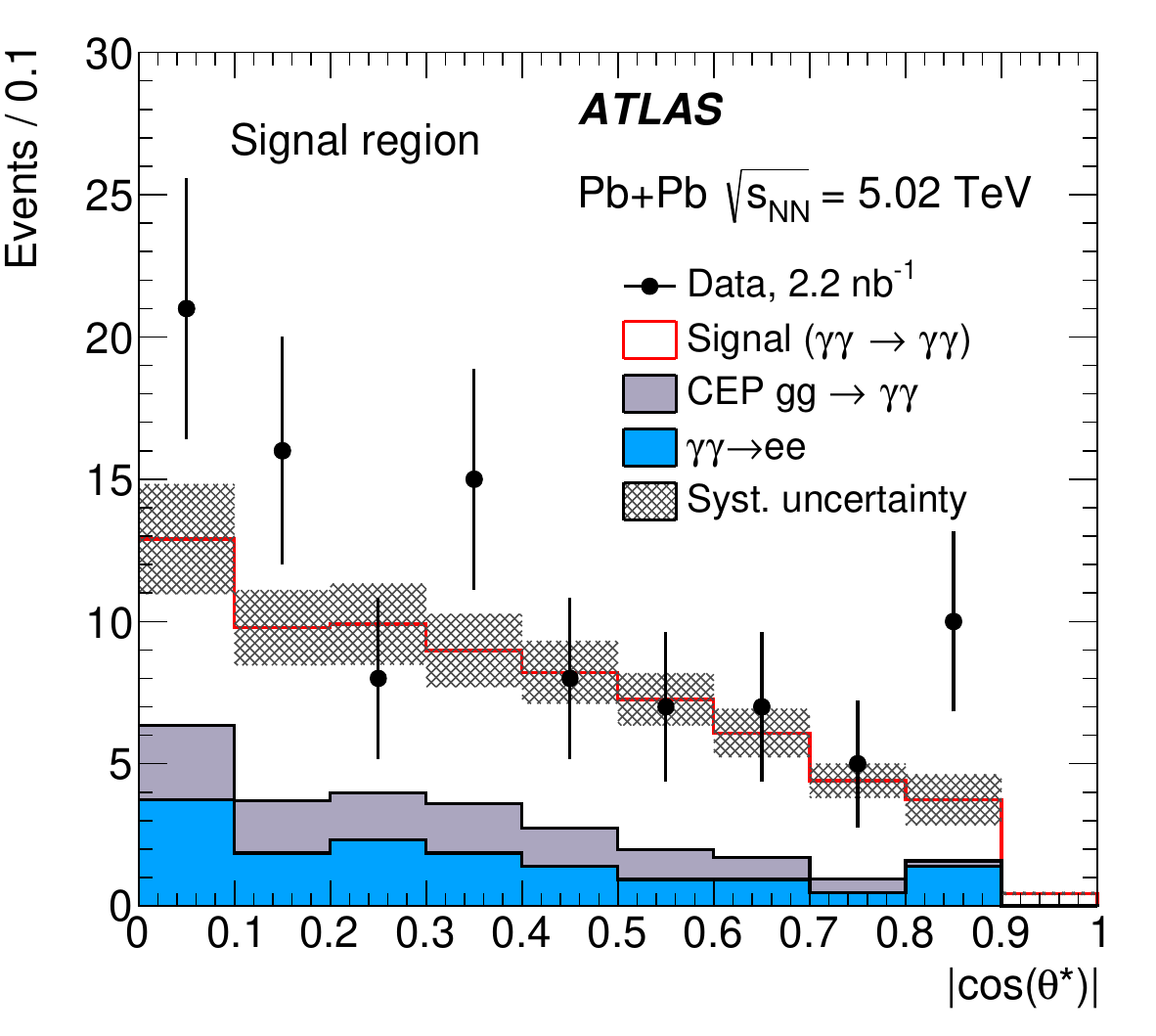}
\includegraphics[width=0.42\textwidth]{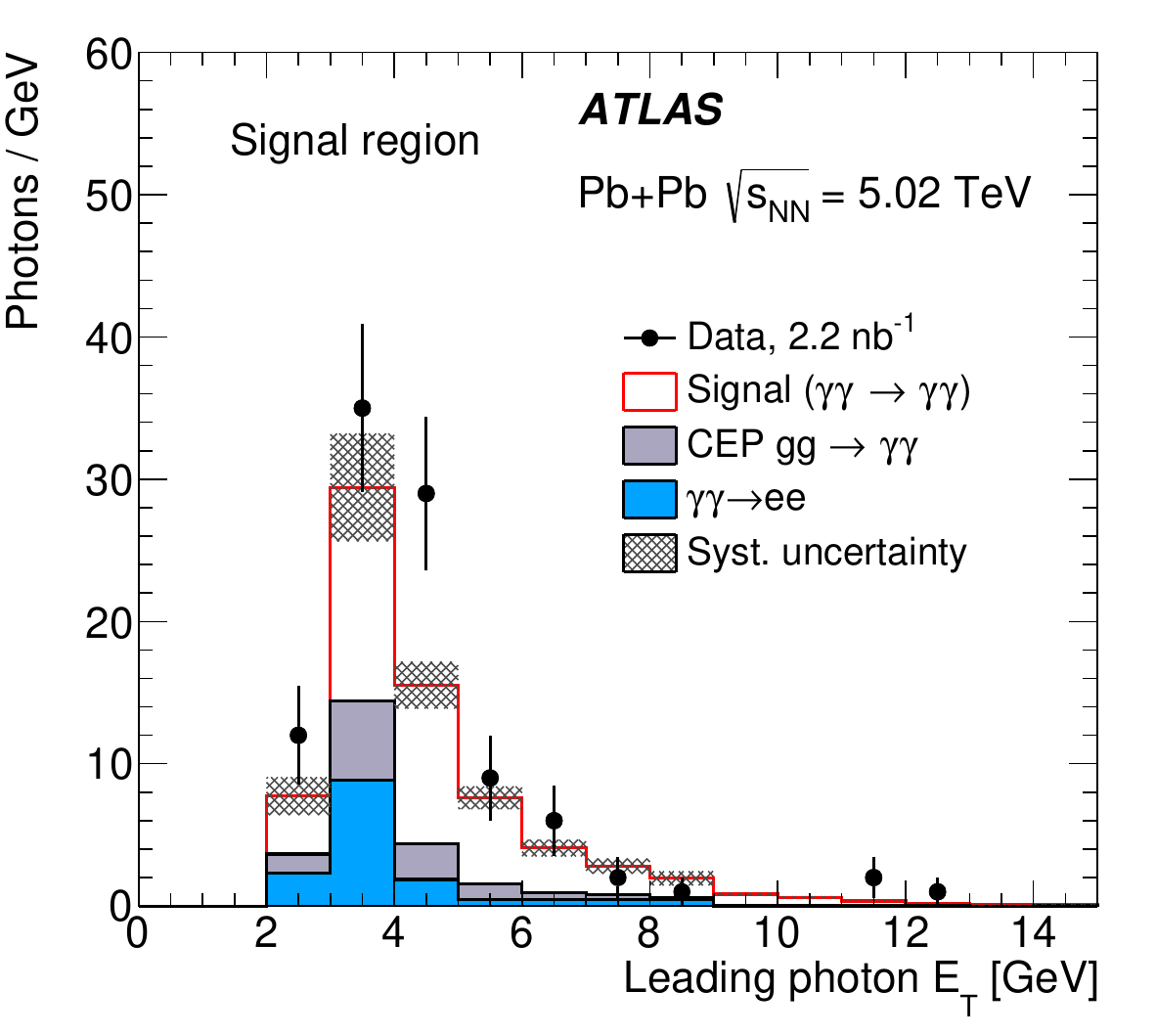}
\includegraphics[width=0.42\textwidth]{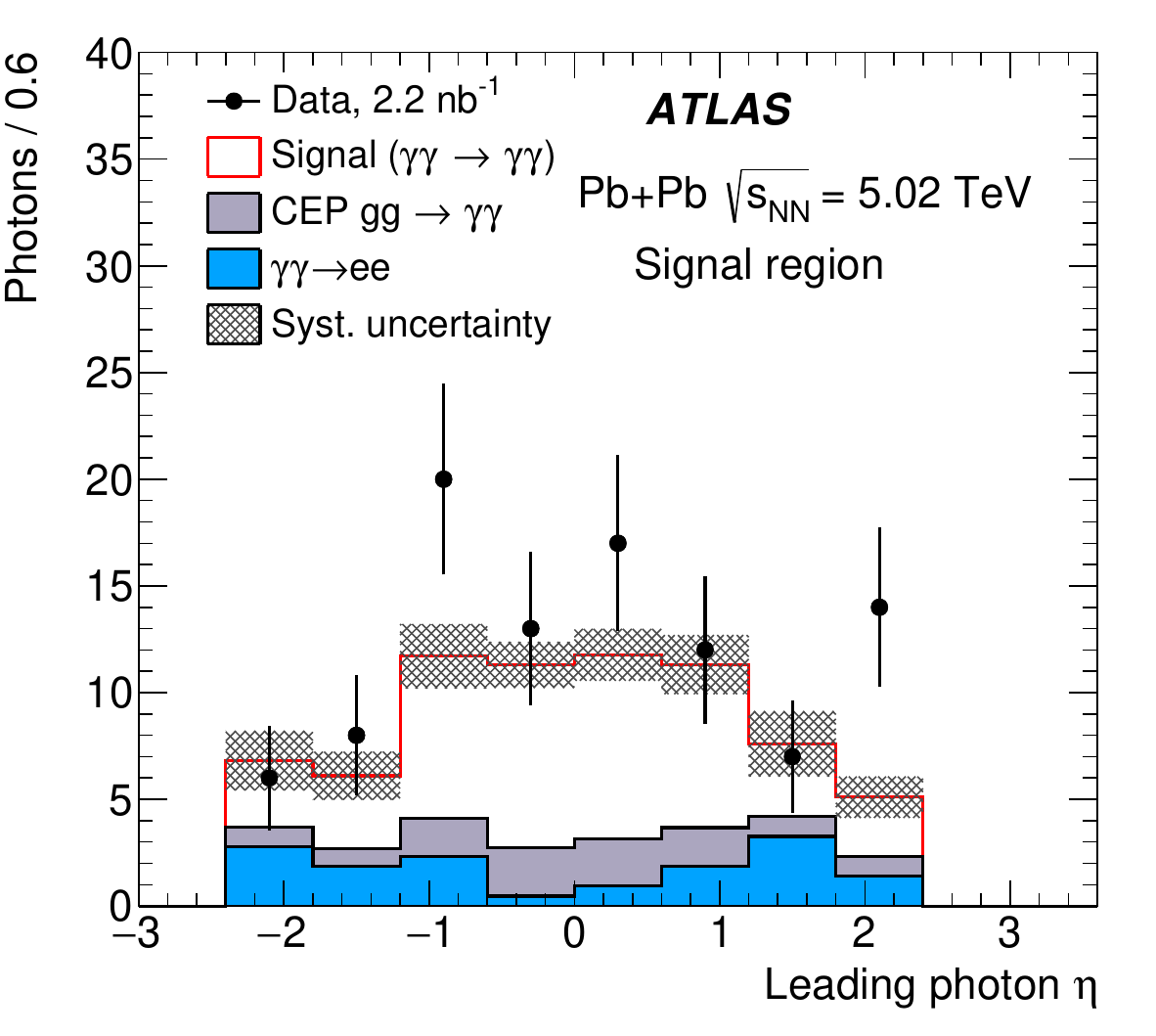}
\caption{Kinematic distributions for \gggg\ event candidates: diphoton invariant mass (top-left), diphoton rapidity (top-right),
diphoton transverse momentum (mid-left), diphoton $|\cos(\theta^*)|$ (mid-right),
leading photon transverse energy (bottom-left) and leading photon pseudorapidity (bottom-right). Data (points) are compared with the sum of signal and background expectations (histograms). The signal prediction is normalised to the same integrated luminosity as the data. Systematic uncertainties in the signal and background processes, excluding that in the luminosity, are shown as shaded bands.}
\label{fig:LbL-control_opt}
\end{center}
\end{figure}

Photon kinematic distributions comparing the selected data
with the sum of expected event yields from simulated signal and background processes in the signal region are shown in Figure~\ref{fig:LbL-control_opt}. In total, 97 events are observed in data where 45 signal events and 27 background events are expected.
This excess of observed events is visible in all distributions shown in Figure~\ref{fig:LbL-control_opt}.

\subsection{\label{sec:fiducial-xsec}Integrated fiducial cross section}
 
The inclusive cross section for the $\gamma\gamma\rightarrow \gamma\gamma$ process is measured in a fiducial phase space, defined by the following requirements on the diphoton final state, reflecting the selection at reconstruction level: both photons have to be within $|\eta|<2.4$ with a transverse momentum of $\pT>2.5~\GeV$. The invariant mass of the diphoton system has to be $m_{\gamma\gamma}>5\,\GeV$ with  transverse momentum of $\pT^{\gamma\gamma}<1~\GeV$. In addition, the photons must fulfil an acoplanarity requirement of $\Aco<0.01$.
 
The integrated fiducial cross section is obtained as follows:
 
\begin{equation}
\label{EQN:CrossSectionFid}
\sigmafid = \frac{N_{\textrm{data}}-N_{\textrm{bkg}}}{C \times \int L \text{d} t}\,,
\end{equation}
 
where $N_{\textrm{data}} = 97$ is the number of selected events in data,  $N_{\textrm{bkg}}=27\pm5$ is the number of background events,  $\int L \text{d} t = 2.22 \pm 0.07~\textrm{nb}^{-1}$ is the integrated luminosity of the data sample and $C=0.263 \pm 0.021$ is
the overall correction factor that accounts for detector efficiencies and resolution effects, and for signal events passing the event selection but originating from outside the fiducial phase space (fiducial corrections).
The $C$ factor is defined as the ratio of the number of reconstructed MC signal events passing the selection to the number of generated MC signal events satisfying the fiducial requirements.
 
The uncertainty in $C$ is estimated by varying the data/MC correction factors within their uncertainties as described in Section~\ref{sec:Systematics}, in particular for the photon reconstruction and PID efficiencies, photon energy scale and resolution and trigger efficiency.
An overview of the various uncertainties in $C$ is given in Table~\ref{tab:AandCFactor}.
 
The uncertainty in $N_{\textrm{bkg}}$ is dominated by the uncertainty
in the \ggee\ background. This has a 6\% impact  on the estimated
integrated fiducial cross section.
 
\begin{table}[t!]
\begin{center}
\begin{tabular}{l |c}
 
\hline
Source of uncertainty			& Detector correction ($C$)	\\
\hline \hline
& $0.263\pm 0.021$ 					\\
\hline
Trigger efficiency 	& 5\%						\\
Photon reco. efficiency 	& 4\%	\\
Photon PID efficiency	& 2\%	\\
Photon energy scale 	& 1\%	\\
Photon energy resolution 	& 2\%   \\
Photon angular resolution 	& 2\%   \\
Alternative signal MC 	& 1\%   \\
Signal MC statistics 	& 1\%   \\
\hline
Total 				& 8\%						\\
\end{tabular}
\caption{The detector correction factor, $C$, and its uncertainties
for the integrated fiducial cross-section measurement.
The second row lists the numerical value of $C$ together with the total uncertainty.
The total uncertainty on $C$ is a quadratic sum of systematic and statistical components.}
\label{tab:AandCFactor}
\end{center}
\end{table}
 
The measured integrated fiducial cross section is $\sigmafid = 120 \pm
17~\textrm{(stat.)} \pm 13~\textrm{(syst.)}\pm 4~\textrm{(lumi.)}$~nb,
which can be compared with the predicted values of $80 \pm 8$~nb from
Ref.~\cite{Klusek-Gawenda:2016euz} and $78 \pm 8$~nb from the SuperChic~v3.0
MC generator~\cite{Harland-Lang:2018iur}. The data-to-theory ratios are $1.50
\pm 0.32$ and $1.54\pm 0.32$, respectively.
 
The theoretical uncertainty in the cross section is primarily due to limited knowledge of the nuclear (EM) form-factors and the related initial photon fluxes. This is extensively studied in Ref.~\cite{Azevedo:2019fyz} and the relevant uncertainty is estimated to be $10\%$ within the fiducial phase space of the measurement.
For masses below 100 \GeV, this uncertainty does not exhibit a dependence
on the diphoton mass.
Higher-order corrections (not included in the calculations) are also part of the theoretical uncertainty and are of the order of 1--3\% in the corresponding invariant mass range~\cite{Bern:2001dg, Klusek-Gawenda:2016nuo}.

\subsection{\label{sec:diff-xsec}Differential fiducial cross sections}
 
Differential fiducial cross sections as a function of  diphoton invariant mass,
diphoton absolute rapidity, average photon transverse momentum
and diphoton $|\cos{\theta^*}|$ are unfolded to particle level in the fiducial phase space
described in the previous section.
 
The differential fiducial cross sections are determined using an iterative Bayesian unfolding method~\cite{DAgostini:1994fjx} with one iteration for all distributions.
The unfolding procedure corrects for bin migrations between particle-
and detector-level distributions due to detector resolution effects,
and applies reconstruction efficiency as well as fiducial corrections.
The reconstruction efficiency corrects for events inside the fiducial region that are not reconstructed in the signal region due to detector inefficiencies; the fiducial corrections take into account events that are reconstructed in the signal region, but originate from outside the fiducial region.
The background contributions are subtracted from data prior to unfolding.

The statistical uncertainty of the data is estimated using 1000 Poisson-distributed pseudo-data sets, constructed by smearing the observed number of events in each bin of the detector-level distribution.
The root mean square of the differences between the resulting unfolded distributions and the unfolded data is taken as the statistical uncertainty in each bin.
 
In the measurement of differential fiducial cross sections, the full
set of experimental systematic uncertainties described in
Section~\ref{sec:Systematics} is considered.
In addition, uncertainties due to the unfolding procedure and the
modelling of the signal process are considered by repeating the
cross-section extraction with modified inputs~\cite{Malaescu:2009dm}.
The distributions are reweighted at generator level to obtain better
agreement between data and simulation after event reconstruction.
The obtained prediction at detector level, which is then similar to
data, is unfolded with the input of the default unfolding and the difference from the
reweighted prediction at generator level is considered as an
uncertainty. The size of this uncertainty is typically below 1\%.
The impact of statistical uncertainties in the signal simulation is estimated using pseudo-data and
is found to be 1--3\%.

\begin{figure}[t!]
\begin{center}
\includegraphics[width=0.43\textwidth]{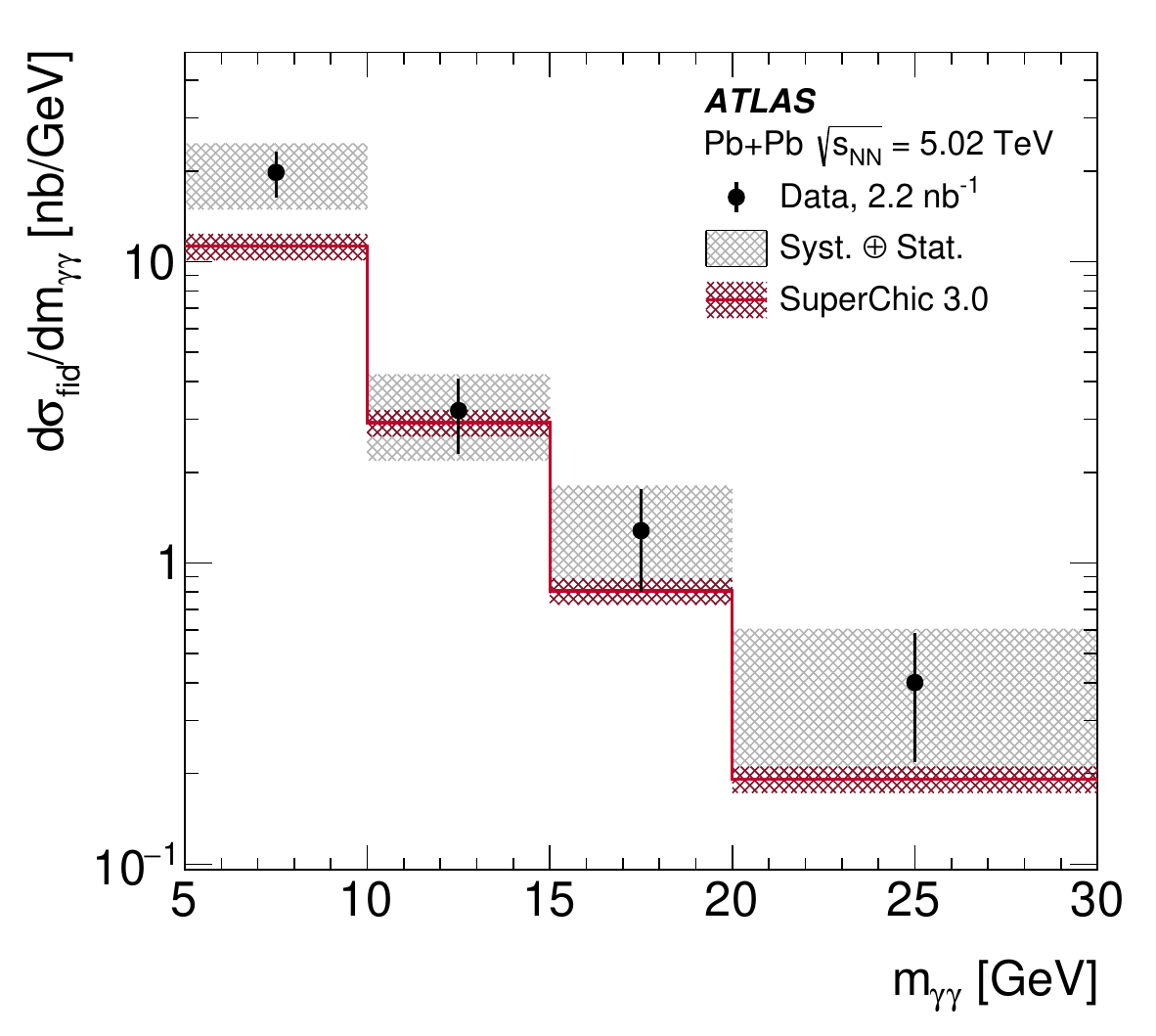}
\includegraphics[width=0.43\textwidth]{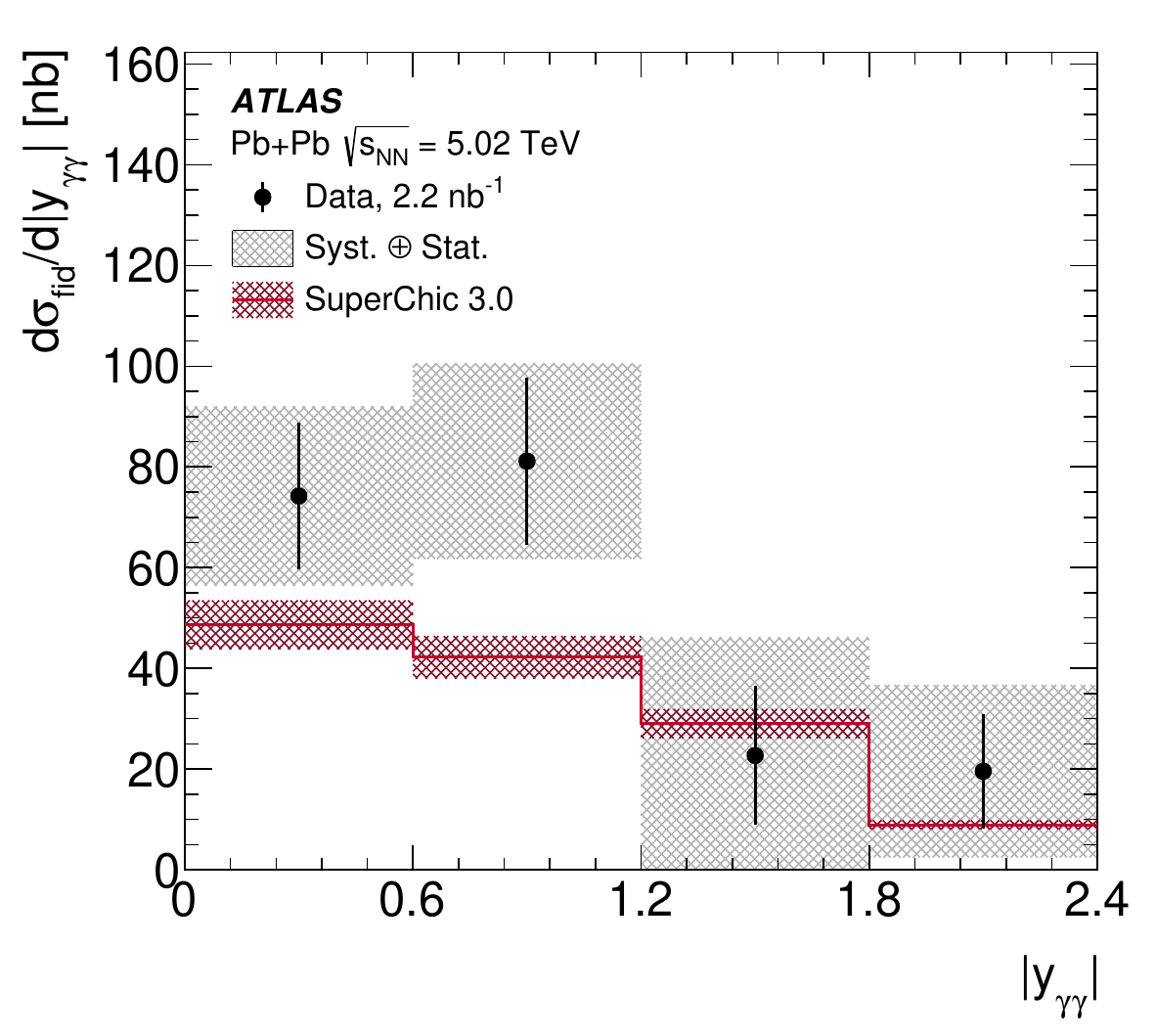}
\includegraphics[width=0.43\textwidth]{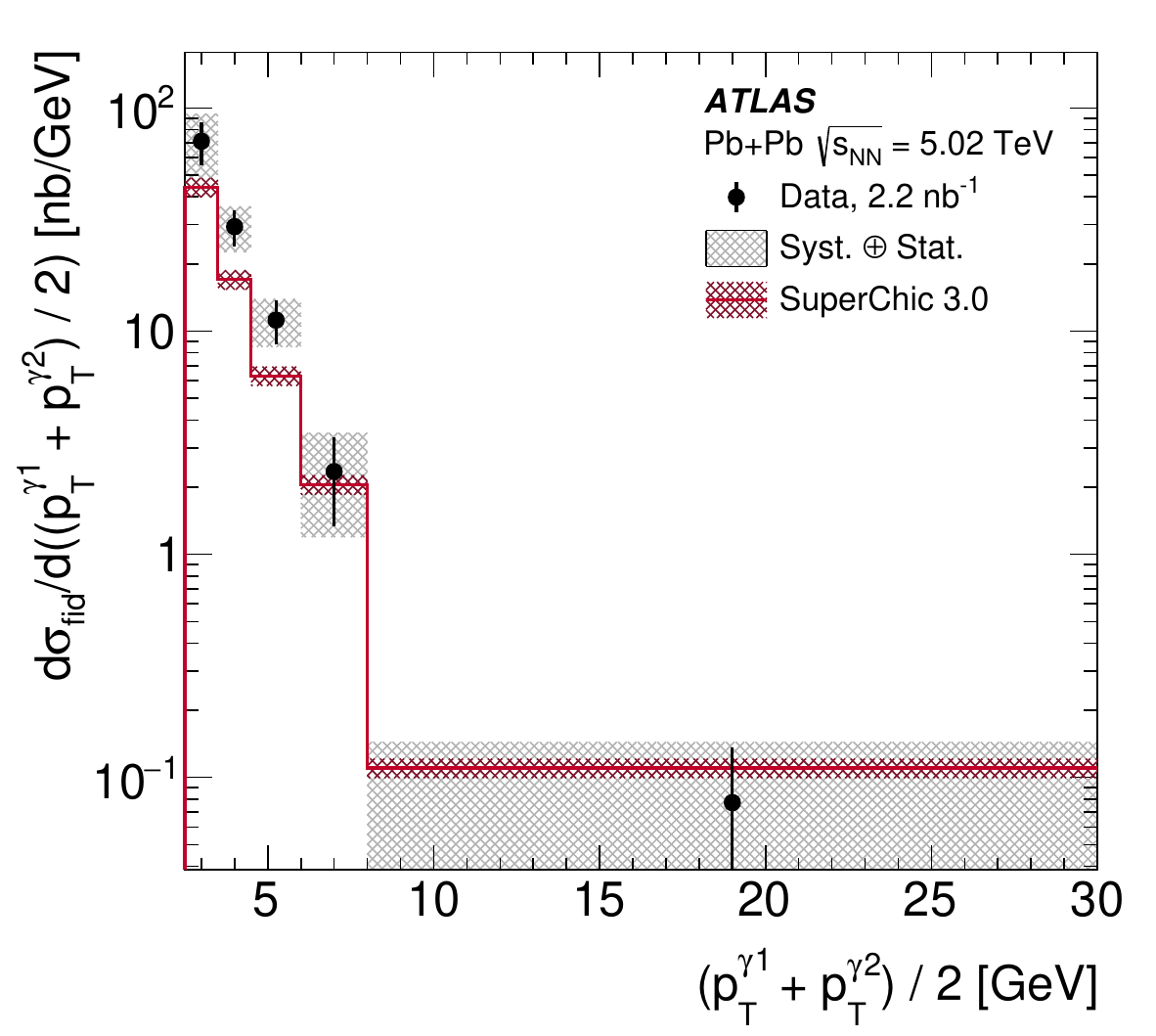}
\includegraphics[width=0.43\textwidth]{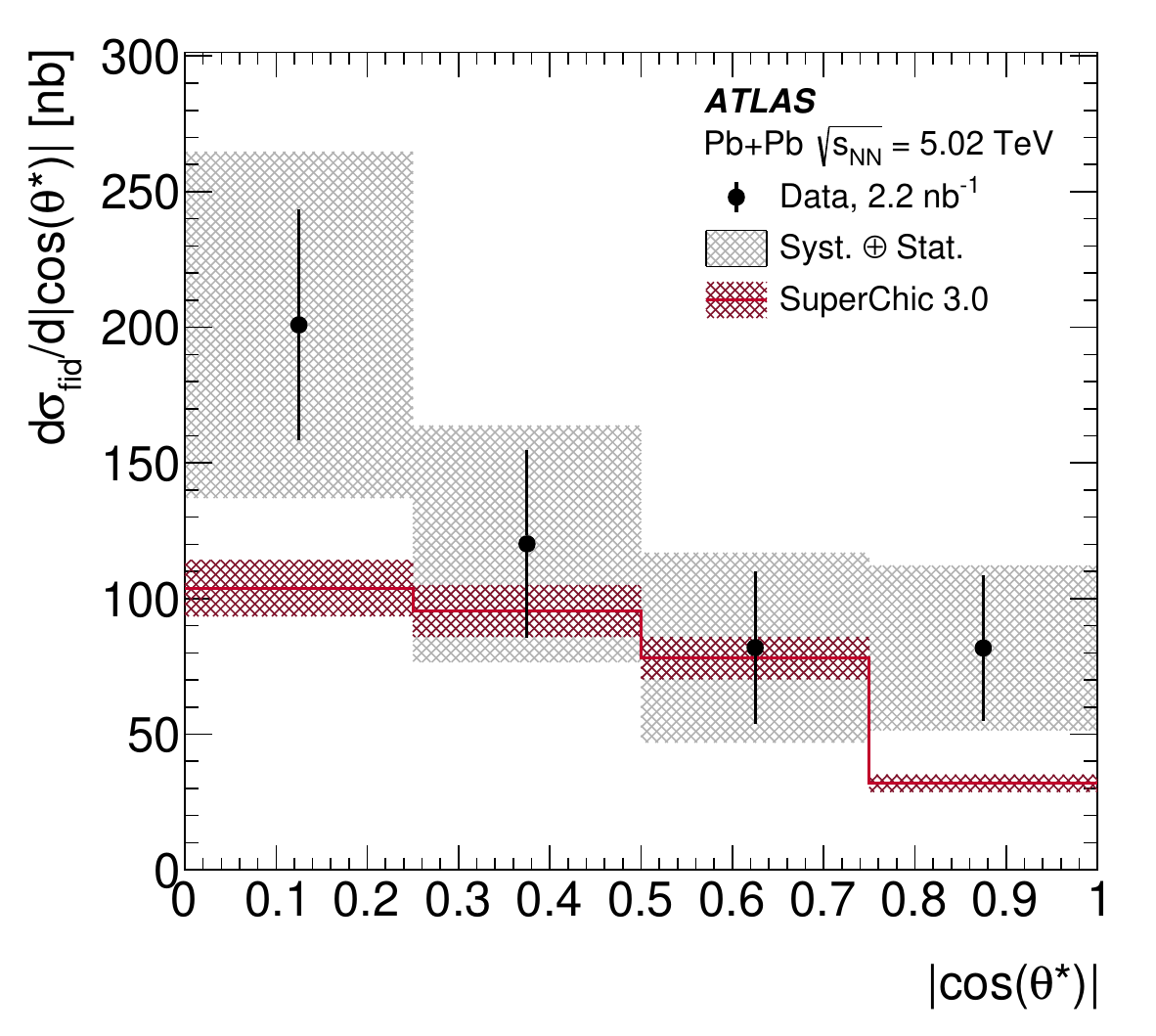}
\caption{Measured differential fiducial cross sections of  \gggg\ production in
Pb+Pb collisions at  $\sqn=5.02$~\TeV\ for four observables (from left
to right and top to bottom): diphoton invariant mass, diphoton absolute rapidity,
average photon transverse momentum and  diphoton $|\cos(\theta^*)|$.
The measured cross-section values are shown as points with error bars giving the statistical uncertainty and grey bands indicating the size of the total uncertainty.
The results are compared with the prediction from the SuperChic~v3.0 MC
generator (solid line) with bands denoting the theoretical uncertainty.
}
\label{fig:LbL-diff}
\end{center}
\end{figure}

The unfolded differential fiducial cross sections are shown in Figure~\ref{fig:LbL-diff}.
They are compared with the predictions from SuperChic~v3.0, which provide a fair description of the data, except for the overall normalisation differences.
For nearly all variables and bins the total uncertainties in the cross-section measurements are dominated by statistical uncertainties, ranging from 25\% to 75\%.
The background systematic uncertainties are large and comparable to statistical uncertainties in some bins (up to 40\%, mainly at high $|y_{\gamma\gamma}|$) due to the limited number of events in the data control regions.
Global $\chi^2$ comparisons are carried out for the shapes of differential distributions.
They do not display any significant differences between predictions and data, with the largest $\chi^2$  per degree of freedom being 4.3/3 when comparing the shape of $|\cos(\theta^*)|$ distribution.
The $\Minvgg$ differential fiducial distribution is measured up to $\Minvgg=30~\GeV$. For $\Minvgg>30~\GeV$, no events are observed in data versus a total expectation of 0.8 events.

The cross sections for all distributions shown in this paper,
including normalised differential fiducial cross sections, are available in HepData~\cite{hepdata}.

\subsection{\label{sec:Alps}Search for ALP production}
 
Any particle coupling directly to photons could be produced in an $s$-channel process in
photon--photon collisions, leading to a resonance peak in the invariant mass
spectrum. One popular candidate for producing a narrow diphoton resonance is an
axion-like particle~(ALP) \cite{Knapen:2016moh}.
The measured diphoton invariant mass spectrum, as shown in Figure~\ref{fig:LbL-control_opt}, is used
to
search for
$\gamma\gamma\rightarrow a \rightarrow\gamma\gamma$ process, where $a$ denotes the ALP.
The LbyL, $\ggee$ and CEP $gg\rightarrow\gamma\gamma$ processes are considered as background.
The contribution from $\ggee$ and CEP $gg\rightarrow\gamma\gamma$ processes is estimated using data-driven techniques as described in Section~\ref{sec:Background}.
The LbyL background is estimated using simulated events generated with SuperChic~v3.0.
These events are normalised to the data yield, after subtracting $\ggee$ and CEP $gg\rightarrow\gamma\gamma$ contributions and excluding the mass search region. To smooth statistical fluctuations in the background shape at high mass, a Crystal Ball function is fitted to the sum of all background contributions, while assigning the fit residuals as additional systematic uncertainty.
 
Events simulated with STARlight~v2.0~\cite{Klein:2016yzr}, which implements the ALP couplings as described in Ref.~\cite{Knapen:2016moh},  for various ALP masses between 5~\GeV\ and 100~\GeV\ are used to build an analytical model of the ALP
signal, interpolating between the simulated mass points.
The efficiency of ALP events to satisfy the selection criteria (outlined in Section~\ref{sec:Selection}) is about 20\% for $\ma=6~\GeV$ and increases up to 45\% for $\ma=12~\GeV$.
An efficiency plateau of about 80\% is reached for an ALP mass around $40~\GeV$.
The diphoton invariant mass resolution for simulated ALP signal ranges from 0.5~\GeV\ at $m_a=6$~\GeV\ to 1.5~\GeV\ at  $m_a=100$~\GeV\ and is dominated by the photon energy resolution.
The impact of the uncertainty on the primary-vertex position has a subdominant effect on the diphoton invariant mass resolution over the full mass range.
 
In every analysis bin a cut-and-count analysis is performed to estimate the
expected numbers of background and signal events.
The bin-width is chosen to include at least 80\% of a reconstructed ALP signal peak within a given bin
and ranges from 2~\GeV\ to 20~\GeV.
To cover the entire mass range, the analysis bins overlap and have an equidistant distance of 1~\GeV\ between the bin centres.
The signal contribution is fitted individually for every bin using a maximum-likelihood fit
implemented in the HistFitter software~\cite{Read:2002hq, Baak:2014wma,Cowan:2010js} which is based on
HistFactory~\cite{Cranmer:2012sba}, RooFit~\cite{RooFit} and RooStats~\cite{Moneta:2010pm}.
 
Since no significant deviation from the background-only hypothesis is observed, the result is then used to estimate the upper
limit on the ALP signal strength~($\mu_\mathrm{CLs}$) at 95\% confidence level~(CL).
The corresponding test-statistic distributions are evaluated using pseudo-experiments.

Experimental systematic uncertainties affecting the ALP signal model originate from the
trigger, photon PID and reconstruction efficiencies, and photon energy scale and resolution.
The systematic uncertainties are evaluated identically to the treatment in the cross-section measurements, described in Section~\ref{sec:Systematics}.
The theoretical uncertainty in the calculated ALP signal cross section is 10\% in the full
mass range, due to the limited knowledge of the initial photon fluxes~\cite{Azevedo:2019fyz}.
This uncertainty is considered uncorrelated with other sources of uncertainty.
 
\begin{figure}[htb]
\begin{center}
\includegraphics[width=0.49\textwidth]{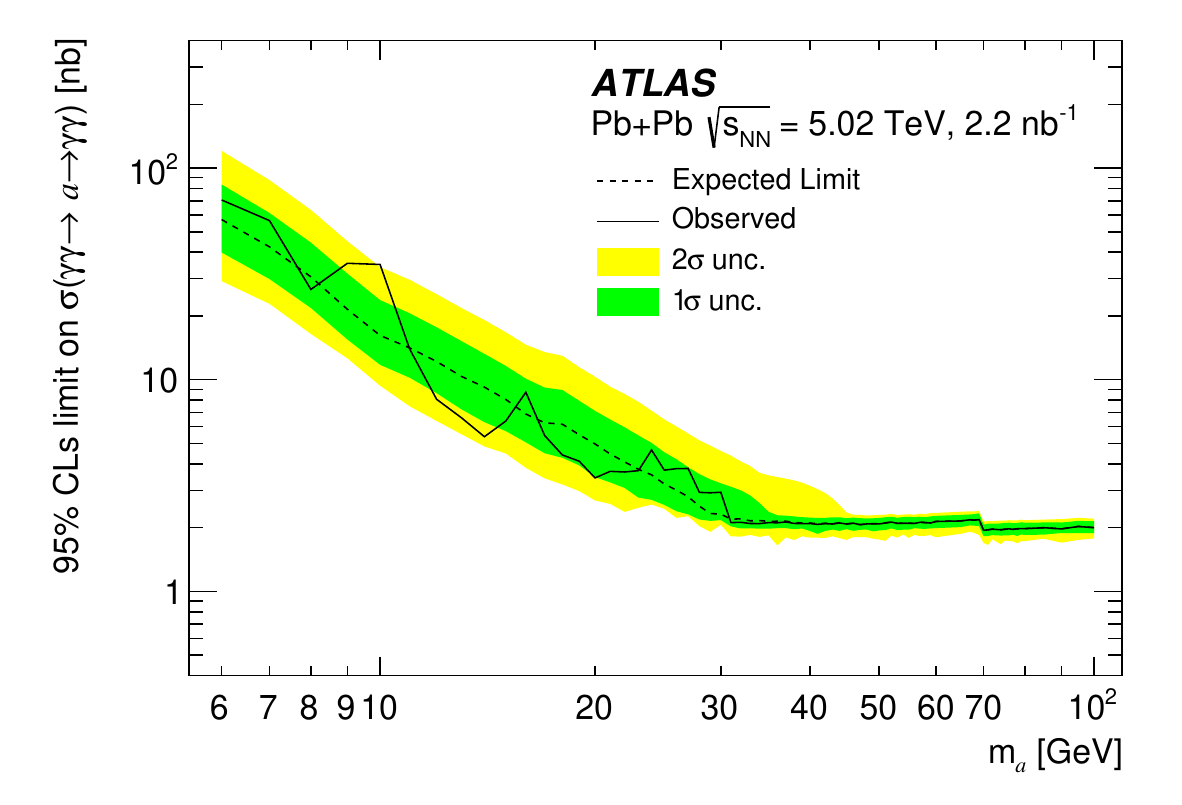}
\includegraphics[width=0.49\textwidth]{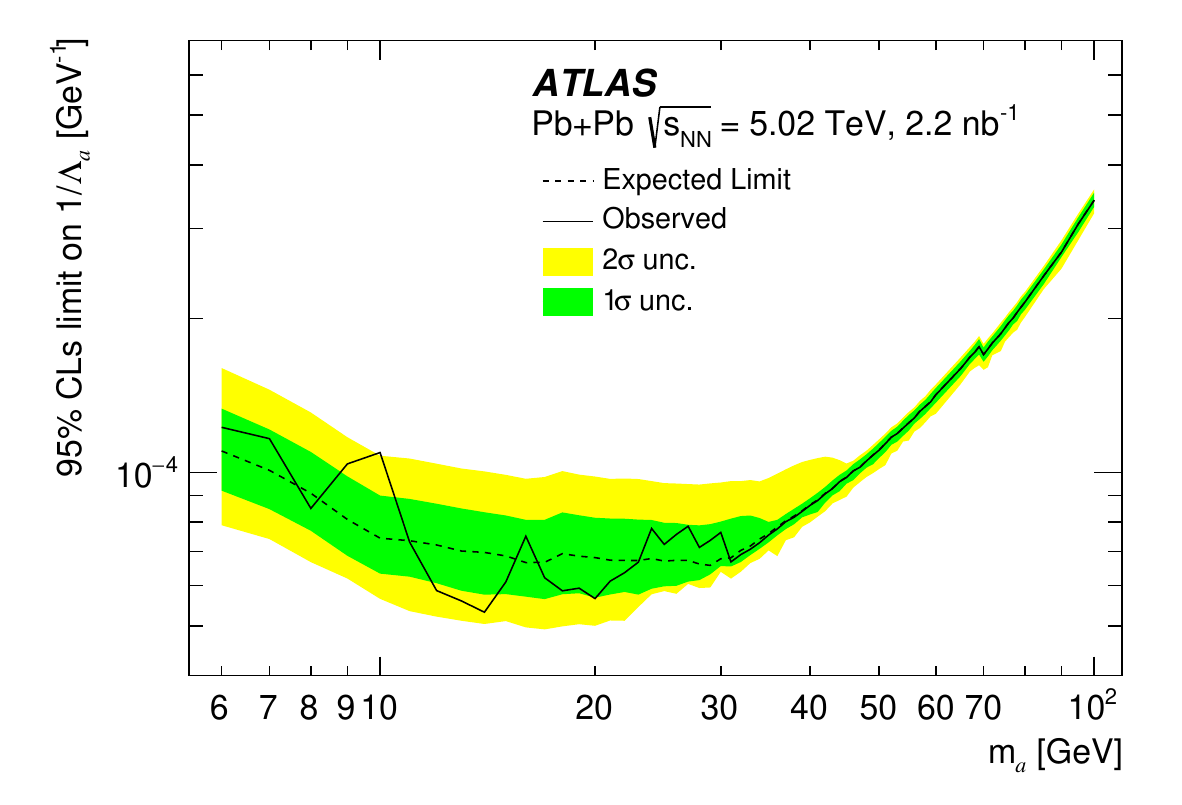}
\caption{
The 95\% CL upper limit on the ALP
cross section $\sigma_{\gamma\gamma\rightarrow a \rightarrow\gamma\gamma}$ (left)  and ALP coupling $1/\Lambda_{a}$ (right) for the $\gamma\gamma\rightarrow a \rightarrow \gamma\gamma$ process as a function of ALP mass \ma.
The observed upper limit is shown as a solid black line and the expected upper limit  is shown by the dashed black line with its $\pm1$ and $\pm2$ standard deviation bands.
The discontinuity at $\ma=70~\GeV$ is caused by the increase of the mass-bin width which brings an increase in signal acceptance.}
\label{fig:ALP_Limits}
\end{center}
\end{figure}

The limits set on the signal strength $\mu_\mathrm{CLs}$ are transformed into
limits on the cross section $\sigma_{\gamma\gamma\rightarrow a \rightarrow\gamma\gamma}^\mathrm{CLs} = \mu_\mathrm{CLs} \cdot \sigma^\mathrm{MC}_{a,\mathrm{gen}}$.
Additionally, limits on the ALP coupling to photons ($1/\Lambda_a^\mathrm{CLs}$)  are calculated from $1/\Lambda_{a}^\mathrm{CLs} = \sqrt{\mu_\mathrm{CLs}} \cdot
1/\Lambda^\mathrm{gen}_{a}$. $\sigma^\mathrm{MC}_{a,\mathrm{gen}}$ and $\Lambda^\mathrm{gen}_{a}$ are the cross section and coupling used in the MC generator.
The observed and expected 95\% CL  limits  on the ALP production cross section  and ALP coupling to photons are presented in Figure~\ref{fig:ALP_Limits}.
The limits set on the cross section $\sigma_{\gamma\gamma\rightarrow a \rightarrow\gamma\gamma}$ for an ALP with a mass of 6--100~\GeV\ range from 70~nb to 2~nb.
The derived constraints on $1/\Lambda_a$ range from 0.3~\TeV$^{-1}$ to 0.06~\TeV$^{-1}$.
The widths of the one- and two-standard-deviation bands of
the expected limit distribution decrease for ALP masses above 30~\GeV. This behaviour is driven by the
change in the background rate, which has a low Poisson mean for high ALP masses. For low ALP masses the
background rate is sufficiently high to populate the $N > 0$ expected background outcomes and raise the +1 and +2-standard-deviation boundaries. The discontinuity at $\ma=70~\GeV$ is caused by the increase of the mass-bin width which brings an increase in signal acceptance.
 
\begin{figure}[htb]
\begin{center}
\includegraphics[width=0.49\textwidth]{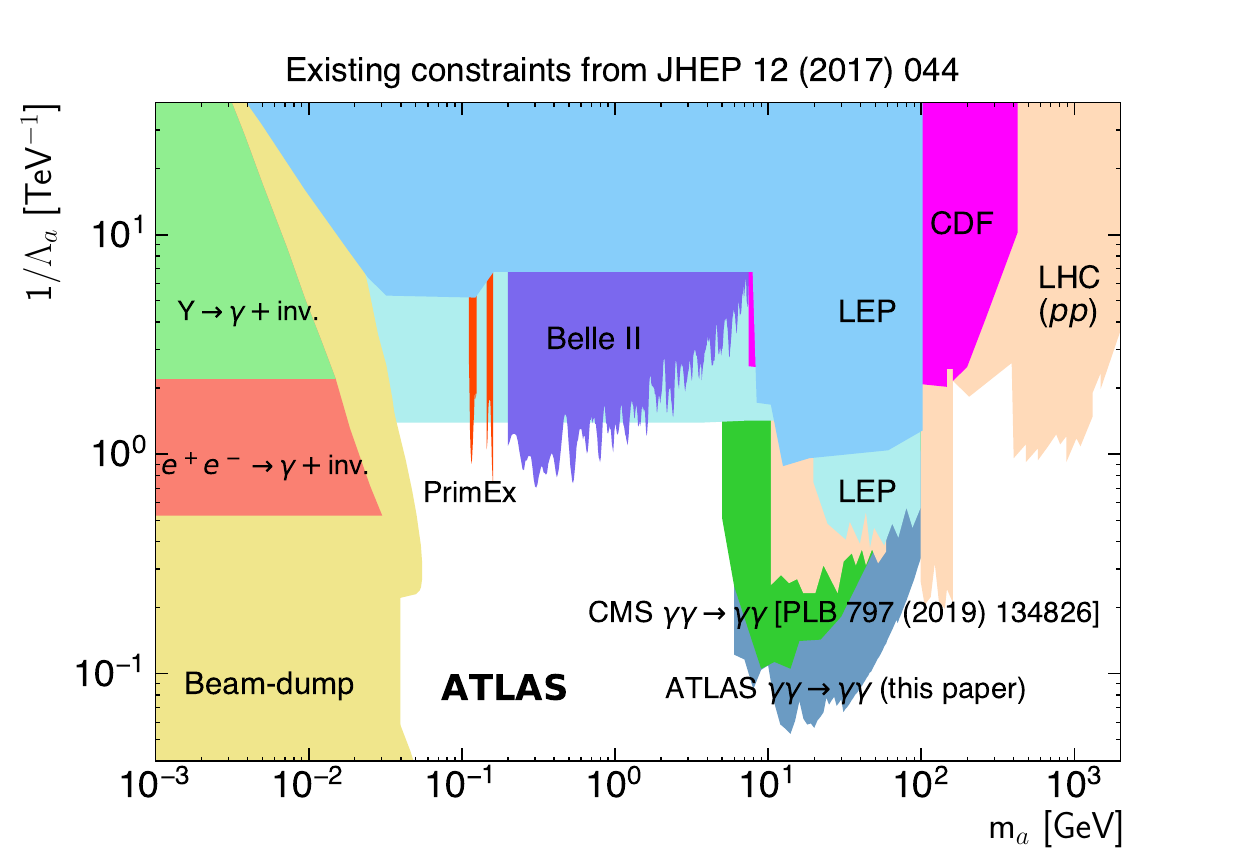}
\includegraphics[width=0.49\textwidth]{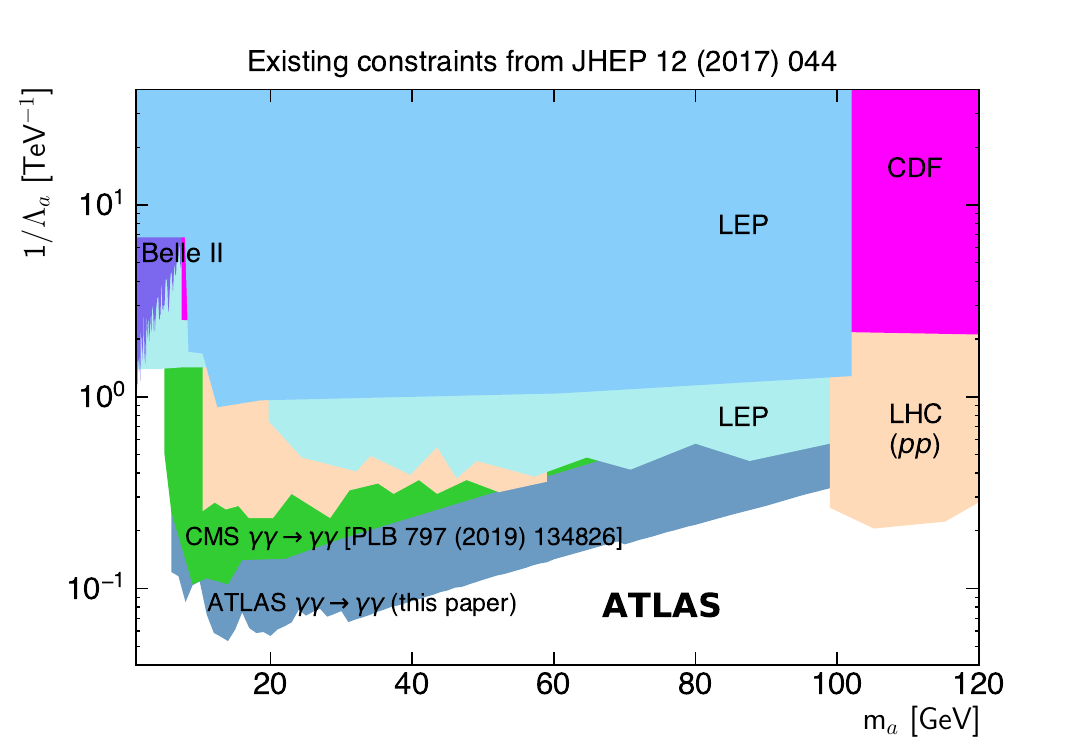}
\caption{Compilation of exclusion limits  at 95\% CL in the ALP--photon coupling ($1/\Lambda_{a}$) versus ALP mass ($\ma$) plane obtained by different experiments. The existing limits, derived from Refs.~\cite{Bauer:2017ris, Sirunyan:2018fhl, Aloni:2019ruo, Banerjee:2020fue, BelleII:2020fag} are compared with the limits extracted from this measurement. The exclusion limits labelled ``LHC ($pp$)'' are based on $pp$ collision data from ATLAS and CMS. All measurements assume a 100\% ALP decay branching fraction into photons. The plot on the right is a zoomed-in version covering the range $1<\ma<120$~\GeV.
}
\label{fig:ALP_Limits_comp}
\end{center}
\end{figure}

Assuming a 100\% ALP decay branching fraction into photons, the derived constraints on the ALP mass and its coupling to photons are compared in Figure~\ref{fig:ALP_Limits_comp} with those obtained from various experiments~\cite{Bauer:2017ris, Sirunyan:2018fhl, Aloni:2019ruo, Banerjee:2020fue, BelleII:2020fag}.
The exclusion limits from this analysis are the strongest so far for the mass range of $6<\ma <100~\GeV$.
 
\clearpage
\section{\label{sec:Conclusion}Conclusions}
 
This paper presents a measurement of the light-by-light scattering process in quasi-real photon interactions from ultra-peripheral \PbPb\ collisions at $\sqn=5.02$~\TeV\ by the ATLAS experiment at the LHC. The measurement is based on the full Run 2 data set corresponding to an integrated luminosity of 2.2~\invnb. After the selection criteria, 97 events are selected in the data while $27\pm5$ background events are expected. The dominant background processes
are estimated using data-driven methods.
 
After background subtraction and corrections for detector effects are applied, the integrated fiducial cross section of the \gggg\ process
is measured to be $\sigmafid = 120 \pm 17~\textrm{(stat.)} \pm 13~\textrm{(syst.)}\pm 4~\textrm{(lumi.)}$ nb.
The data-to-theory ratios are $1.50 \pm 0.32$ and $1.54\pm 0.32$ for predictions from Ref.~\cite{Klusek-Gawenda:2016euz} and from the SuperChic~v3.0 MC generator, respectively.
Differential fiducial cross sections are measured as a function of several properties of the final-state photons and are compared with Standard Model theory predictions for light-by-light scattering.
All measured cross sections are consistent within 2 standard deviations with the predictions.
The measurement precision is limited in all kinematic regions by statistical uncertainties.
 
The measured diphoton invariant mass distribution is used to search for axion-like particles and set new exclusion limits on their production in the $\textrm{Pb+Pb}\,(\gamma\gamma)\rightarrow \textrm{Pb}^{(\ast)}\textrm{+}\textrm{Pb}^{(\ast)}\,\gamma\gamma$ reaction. Integrated cross sections above 2 to 70 nb are excluded at the 95\% CL, depending on the diphoton invariant mass in the range 6--100~\GeV. These results provide, to this date and within the aforementioned mass range, the most stringent constraints in the search for ALP signals.


\section*{Acknowledgements}

We thank CERN for the very successful operation of the LHC, as well as the
support staff from our institutions without whom ATLAS could not be
operated efficiently.
 
We acknowledge the support of ANPCyT, Argentina; YerPhI, Armenia; ARC, Australia; BMWFW and FWF, Austria; ANAS, Azerbaijan; SSTC, Belarus; CNPq and FAPESP, Brazil; NSERC, NRC and CFI, Canada; CERN; ANID, Chile; CAS, MOST and NSFC, China; COLCIENCIAS, Colombia; MSMT CR, MPO CR and VSC CR, Czech Republic; DNRF and DNSRC, Denmark; IN2P3-CNRS and CEA-DRF/IRFU, France; SRNSFG, Georgia; BMBF, HGF and MPG, Germany; GSRT, Greece; RGC and Hong Kong SAR, China; ISF and Benoziyo Center, Israel; INFN, Italy; MEXT and JSPS, Japan; CNRST, Morocco; NWO, Netherlands; RCN, Norway; MNiSW and NCN, Poland; FCT, Portugal; MNE/IFA, Romania; JINR; MES of Russia and NRC KI, Russian Federation; MESTD, Serbia; MSSR, Slovakia; ARRS and MIZ\v{S}, Slovenia; DST/NRF, South Africa; MICINN, Spain; SRC and Wallenberg Foundation, Sweden; SERI, SNSF and Cantons of Bern and Geneva, Switzerland; MOST, Taiwan; TAEK, Turkey; STFC, United Kingdom; DOE and NSF, United States of America. In addition, individual groups and members have received support from BCKDF, CANARIE, Compute Canada, CRC and IVADO, Canada; Beijing Municipal Science \& Technology Commission, China; COST, ERC, ERDF, Horizon 2020 and Marie Sk{\l}odowska-Curie Actions, European Union; Investissements d'Avenir Labex, Investissements d'Avenir Idex and ANR, France; DFG and AvH Foundation, Germany; Herakleitos, Thales and Aristeia programmes co-financed by EU-ESF and the Greek NSRF, Greece; BSF-NSF and GIF, Israel; La Caixa Banking Foundation, CERCA Programme Generalitat de Catalunya and PROMETEO and GenT Programmes Generalitat Valenciana, Spain; G\"{o}ran Gustafssons Stiftelse, Sweden; The Royal Society and Leverhulme Trust, United Kingdom.
 
The crucial computing support from all WLCG partners is acknowledged gratefully, in particular from CERN, the ATLAS Tier-1 facilities at TRIUMF (Canada), NDGF (Denmark, Norway, Sweden), CC-IN2P3 (France), KIT/GridKA (Germany), INFN-CNAF (Italy), NL-T1 (Netherlands), PIC (Spain), ASGC (Taiwan), RAL (UK) and BNL (USA), the Tier-2 facilities worldwide and large non-WLCG resource providers. Major contributors of computing resources are listed in Ref.~\cite{ATL-SOFT-PUB-2020-001}.
 

\printbibliography

\clearpage \input{atlas_authlist}
\end{document}

%% file: atlas_authlist.tex
 
\begin{flushleft}
\hypersetup{urlcolor=black}
{\Large The ATLAS Collaboration}

\bigskip

\AtlasOrcid[0000-0002-6665-4934]{G.~Aad}$^\textrm{\scriptsize 102}$,    
\AtlasOrcid[0000-0002-5888-2734]{B.~Abbott}$^\textrm{\scriptsize 128}$,    
\AtlasOrcid{D.C.~Abbott}$^\textrm{\scriptsize 103}$,    
\AtlasOrcid[0000-0002-2788-3822]{A.~Abed~Abud}$^\textrm{\scriptsize 36}$,    
\AtlasOrcid[0000-0002-1002-1652]{K.~Abeling}$^\textrm{\scriptsize 53}$,    
\AtlasOrcid[0000-0002-2987-4006]{D.K.~Abhayasinghe}$^\textrm{\scriptsize 94}$,    
\AtlasOrcid[0000-0002-8496-9294]{S.H.~Abidi}$^\textrm{\scriptsize 167}$,    
\AtlasOrcid[0000-0002-8279-9324]{O.S.~AbouZeid}$^\textrm{\scriptsize 40}$,    
\AtlasOrcid{N.L.~Abraham}$^\textrm{\scriptsize 156}$,    
\AtlasOrcid[0000-0001-5329-6640]{H.~Abramowicz}$^\textrm{\scriptsize 161}$
% ...
\bigskip
\\

% ...
$^{36}$CERN, Geneva; Switzerland.\\
% ...
$^{40}$Niels Bohr Institute, University of Copenhagen, Copenhagen; Denmark.\\
% ...
$^{53}$II. Physikalisches Institut, Georg-August-Universit\"{a}t G\"ottingen, G\"ottingen; Germany.\\
% ...
$^{94}$Department of Physics, Royal Holloway University of London, Egham; United Kingdom.\\
% ...
$^{102}$CPPM, Aix-Marseille Universit\'e, CNRS/IN2P3, Marseille; France.\\
$^{103}$Department of Physics, University of Massachusetts, Amherst MA; United States of America.\\
% ...
$^{128}$Homer L. Dodge Department of Physics and Astronomy, University of Oklahoma, Norman OK; United States of America.\\
% ...
$^{156}$Department of Physics and Astronomy, University of Sussex, Brighton; United Kingdom.\\
% ...
$^{161}$Raymond and Beverly Sackler School of Physics and Astronomy, Tel Aviv University, Tel Aviv; Israel.\\
% ...
$^{167}$Department of Physics, University of Toronto, Toronto ON; Canada.\\
% ...

\end{flushleft}
